\documentclass[a4paper,10pt]{article}
\usepackage{graphicx}
\usepackage{color}
\usepackage{fancyhdr}
\usepackage{rotating}
\usepackage{float}
\usepackage{pdflscape}
\begin{document}
\sloppy

\title{X-ray and neutron scattering on disordered nanosize clusters: a case study of lead-zirconate-titanate solid solutions}
\author{Johannes Frantti\footnote{Email: johannes.frantti@fre.fi} and Yukari Fujioka, \\Finnish Research and Engineering, \\Jaalaranta 9 B 42, 00180 Helsinki, Finland}
\maketitle
\tableofcontents
\newpage
\begin{abstract}
Defects and frequently used defect models of solids are reviewed. Signatures for identifying the disorder from x-ray and neutron 
scattering data are given. 
To give illustrative examples how technologically important defects contribute to x-ray and neutron scattering numerical 
method able to treat non-periodical solids possessing several simultaneous defect types is given for simulating scattering in nanosize 
disordered clusters. The approach takes particle size, shape, and defects into account and isolates element specific signals. 
As a case study a statistical approximation model for lead-zirconate titanate [Pb(Zr$_x$Ti$_{1-x}$)O$_3$, PZT] is introduced. PZT is a material 
possessing several defect types, including substitutional, displacement and surface defects. 
Spatial composition variation is taken into account by introducing a model in which the edge lengths of each cell 
depend on the distribution of Zr and Ti ions in the cluster. Spatially varying edge lengths and angles 
is referred to as microstrain. 
The model is applied to compute the scattering from ellipsoid shaped PZT clusters and to simulate the structural 
changes as a function of average composition. Two-phase co-existence range, the so called morphotropic phase boundary
composition is given correctly. 
The composition at which the rhombohedral and tetragonal cells are equally abundant was $x\approx 0.51$. 
Selected x-ray and neutron Bragg reflection intensities and line shapes were simulated. 
Examples of the effect of size and shape of the scattering clusters on diffraction patterns are given and the particle dimensions,
computed through Scherrer equation, are compared with the exact cluster dimensions. Scattering from two types of 
180$^{\circ}$ domains in spherical particles, one type assigned to Ti-rich PZT and the second to the MPB and Zr-rich PZT, is computed.
We show how the method can be used for modelling polarization reversal.
\end{abstract}

\section{Introduction}
The paper is organized as follows. Section \ref{xsns} summarizes the basic concepts of x-ray and neutron scattering, sections \ref{models} and \ref{pdf} review the models of disordered 
materials and summarize the pair distribution function method, respectively, and finally a brief review of the crystal structures of lead-zirconate-titanate [Pb(Zr$_x$Ti$_{1-x}$)O$_3$, PZT] 
is given in section \ref{pzt_structures}. PZT was chosen 
as an example material due to its wide use in applications, which are often based on the controlled use of defects.
Section \ref{PZT_case} describes the numerical method applied 
in the present study. Section \ref{cluster_section} describes the method after which selected case studies on PZT clusters are given in sections  \ref{spherical_section}, \ref{ellipsoid_section} and \ref{DomainSection}. 
\subsection{\label{xsns}X-ray and neutron diffraction.}
\begin{figure}[h!]
\begin{center}
\includegraphics[angle=0,width=6cm]{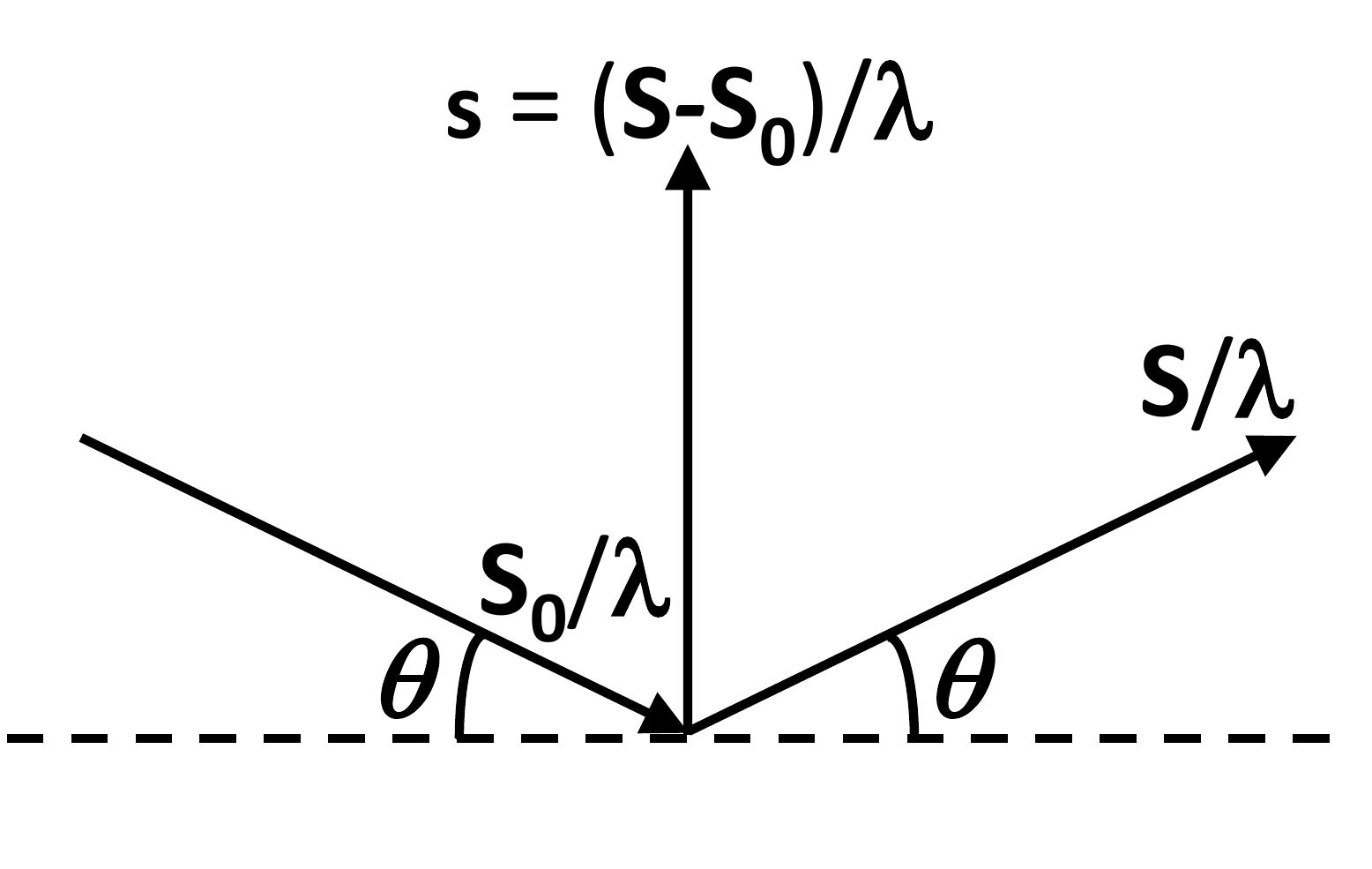}
\caption{\label{Scattering}Geometrical description of the scattering vector $\mathbf{s}$.}
\end{center}
\end{figure}

Diffraction techniques, notably x-ray diffraction, are the most commonly applied materials characterization methods. 
The theory of x-ray scattering and diffraction is well established and below key principles (see, e.g., ref. \cite{Guinier}) are given. The far-field amplitude $A(\mathbf{s})$ is given by Eq. (\ref{Amplitude}) 
\begin{center}
\begin{equation}\label{Amplitude}
A(\mathbf{s})=\sum_n^Nf_ne^{2\pi i \mathbf{s}\cdot\mathbf{x}_n},
\end{equation}
\end{center} 
where $N$ is the number of atoms in the scattering volume, $f_n$ and $\mathbf{x}_n$ are the scattering amplitude and the position vector of the atom n, respectively, and $\mathbf{s}$ is the scattering vector, depicted in Fig. \ref{Scattering}. 
By denoting the unit propagation vectors of the incoming and scattered radiation of wavelength $\lambda$  by 
$\mathbf{S}_0$ and $\mathbf{S}$ respectively, $\mathbf{s}$ is $(\mathbf{S}-\mathbf{S}_0)/\lambda$. In crystals the intensity maxima 
correspond to the reciprocal lattice points $\mathbf{s}=h\mathbf{a}^*+k\mathbf{b}^*+l\mathbf{c}^*$, where 
$\mathbf{a}^*$, $\mathbf{b}^*$ and $\mathbf{c}^*$ are the reciprocal lattice vectors and $h$, $k$ and $l$ are integers.
Eq. (\ref{Amplitude}) can be generalized by replacing the discrete atomic densities by a continuous electron density $\rho(\mathbf{x})$:
\begin{center}
\begin{equation}\label{E_Density}
A(\mathbf{s})=\int \rho(\mathbf{x}) e^{-2\pi i \mathbf{s}\cdot\mathbf{x}}dv_x,
\end{equation}
\end{center} 
so that $A(\mathbf{s})$ is the Fourier transform of $\rho(\mathbf{x})$. Inversely, electron density is given by Eq. (\ref{Inverse})
\begin{center}
\begin{equation}\label{Inverse}
\rho(\mathbf{x})=\int  A(\mathbf{s}) e^{2\pi i \mathbf{s}\cdot\mathbf{x}}dv_s.
\end{equation}
\end{center}
The scattered intensity $I_N(\mathbf{s})$ is given by Eq. (\ref{Intensity})
\begin{center}
\begin{equation}\label{Intensity}
I_N(\mathbf{s})=\vert  A(\mathbf{s}) \vert^2
\end{equation}
\end{center}
and, in terms of the convolution obtained by substituting Eq. (\ref{Amplitude}) to  Eq. (\ref{Intensity}), $I_N=\int\int\rho(\mathbf{u})\rho(\mathbf{x+u})e^{-2\pi i \mathbf{s}\cdot\mathbf{x}}dv_udv_x$ which 
reads that the Fourier-transform of the autocorrelation function (the Patterson function) equals to the intensity.
In the case of an infinite crystal the electron density can be expressed as a convolution between the function representing the electron density inside a unit cell and a series of Dirac functions representing the 
crystal lattice. Fourier-transform of the convolution results in the structure factor $F_{hkl}$ and further gives the well-known result according to which the intensity of the reflection $hkl$ is $\vert F_{hkl} \vert^2$.
Section \ref{PZT_case} focuses on the nanosize clusters which possess short-range order but lack translational symmetry. The nanosize itself does not remove the translational symmetry: 
we consider cases in which each 'unit cell' has own 'lattice parameters' (i.e., the assumption of translational symmetry is abandoned).

If the absorption, extinction, thermal displacements, angle dependent polarization corrections and 
instrument related factors are not considered, the elastic scattering intensity can readily be computed from Eq. (\ref{IN}) once the atomic scale structure is known:
\begin{center}
\begin{equation}\label{IN}
I_N(\mathbf{s})=\sum_{n,n'}^Nf_nf_{n'}\cos[2\pi\mathbf{s}\cdot(\mathbf{x}_n-\mathbf{x}_{n'})].
\end{equation}
\end{center}
Neutron diffraction intensity is modelled similarly, the scattering amplitudes are replaced by nuclei $n$ specific neutron scattering lengths $b_0(n)$. 
Since x-rays interact with electrons and neutrons interact with nuclei the techniques are in many ways complementary. 
In contrast to x-rays, neutrons frequently scatter strongly from light nuclei, such as oxygen and hydrogen, which is essentially 
why neutrons suit for the determination of light elements positions. The neutron scattering lengths of different isotopes are 
often quite different, even possessing different signs, which has made isotope substitution a technique for pinpointing structural details.
Since neutrons possess a magnetic moment they interact with the magnetic moments of electrons, the cross section being the same order of 
magnitude as the neutron-nuclear interaction. 
Correspondingly, neutron scattering is the most common method for determining magnetic ordering. 
Magnetic form factors used in the calculations of the cross sections for magnetic scattering of neutrons are given in 
terms of sums of exponential functions whose coefficients are tabulated in ref. \cite{ITCC}. The form factors decay with increasing 
$\mathbf{s}$ so that often the minimum $d$-spacing included in the magnetic scattering model is $\approx 1$\AA. The form factors 
depend on the valence state of the ions and thus it is frequently necessary to complement neutron scattering data by alternative measurements
to clarify the valence state(s) of the ions.
Since the scattering power of atoms for neutrons is not $Q$-dependent ($Q=4\pi \sin\theta/\lambda$), in contrast to x-rays where the atomic 
scattering factors fall away rapidly at high-$Q$, strong diffuse scattering can be observed well beyond the $Q$-range where Bragg peaks 
occur \cite{Welberry}.

\subsection{\label{models}Common Models for Disordered Systems}
Computation of diffraction intensities is straightforward once the positions of atoms are given (see, for instance refs. \cite{Guinier} and \cite{Woolfson}). 
The most challenging  problem is to find the atomic scale structure (structural model) corresponding to the measured intensity. Numerous 
recipes to solve the problem have been developed, for instance see  refs. \cite{Woolfson} and \cite{Massa}. Space group symmetry 
determination from the Laue symmetry and the reflection conditions, as obtained from the diffraction patterns, is 
given in ref. \cite{ITCA}. At the final stage of the structure solving task one introduces a model which is refined 
by adjusting model parameters so that the difference between the computed intensity and the measured intensity is minimized. 
A well-known refinement technique for powders is the Rietveld method \cite{Rietveld_1969,Young}. 
In the case of known average symmetry, equation (\ref{IN}) is not applied directly, but one computes the 
squared value of the absolute value of the structure factor, $\vert F_{hkl}\vert^2$ of the Bragg reflection $hkl$. 
$\vert F_{hkl}\vert^2$, however, tells 
nothing about the linewidths or the shape of the profile function of the reflection. Numerous line shapes were derived for different instruments and sometimes the choice of the profile function for describing the sample and instrument 
originating line broadening is challenging. The broadening parameters are refined 
as a part of the structural model. More recently, instrument related line broadening has also been computed from 
the known diffractometer properties as summarized in ref. \cite{Birkholz}. In principle, direct application of
 Eq. (\ref{IN}) gives the \emph{sample} contribution to the lineshape. 

Direct application of equation (\ref{IN}) is  a heavy computational task even for relatively small atom clusters.
Generally, the problem is challenging once the material lacks periodicity at least in one dimension as even a straightforward 
computation of the scattered intensity from a collection of atoms with known positions becomes computationally 
formidable task (see, however, Model \#5 in section \ref{models}). Thus, to make structure solving or even refinement possible approximations are required.  
The most complex structures require tailored solutions as they do not possess the high symmetry required by the commonly available programs.

\begin{figure}[h!]
\begin{center}
\includegraphics[angle=0,width=12cm]{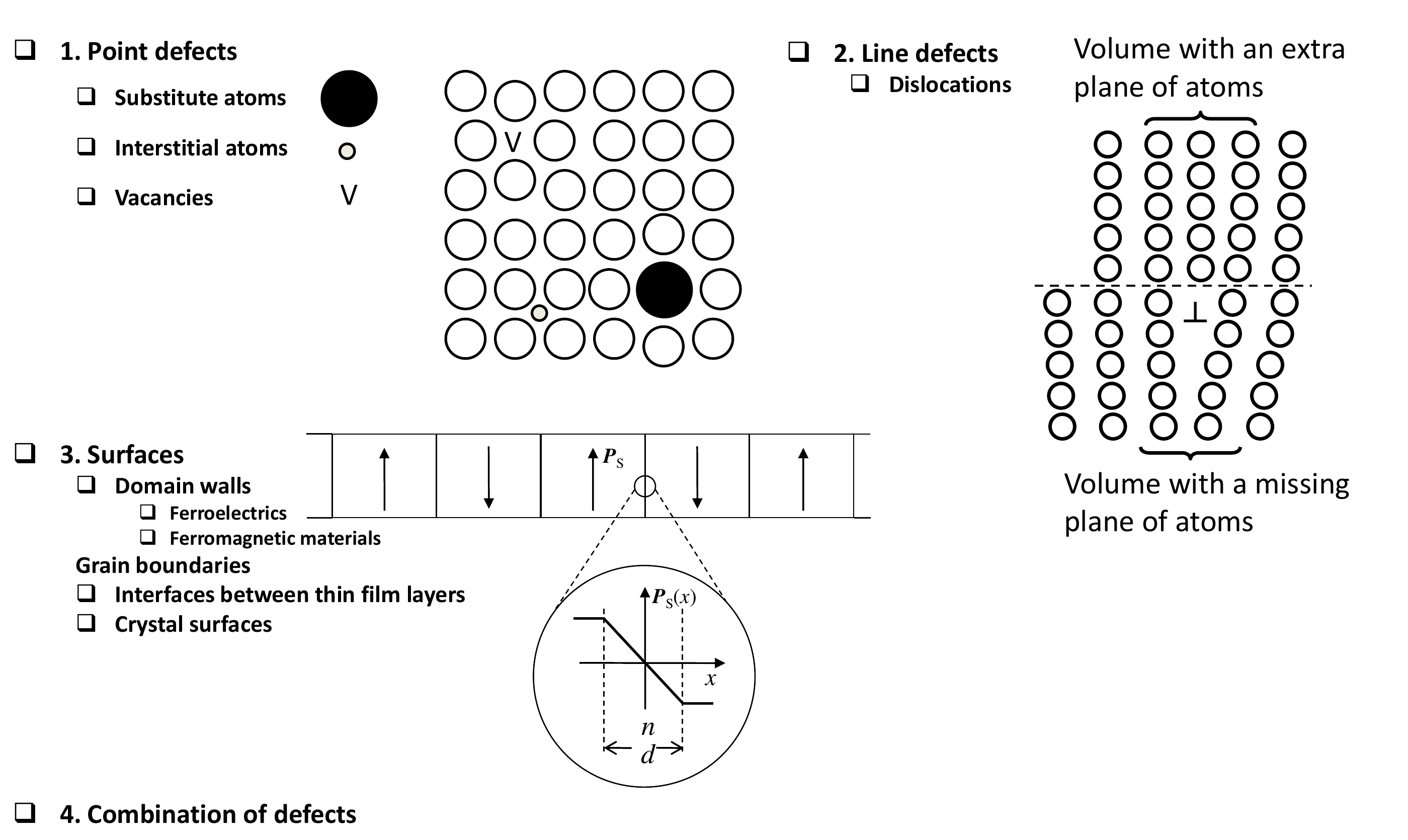}
\caption{\label{FigDefects}Examples of point, line and two-dimensional crystal defects, which frequently occur together. Point defects include 
substitute atoms, vacancies and interstitial atoms and are accompanied by a local strain field. Edge dislocations are common in thin film 
multilayer structures and occur when the mismatch between the thin film layer and substrate is so large that strain is energetically 
unfavourable. Domains in ferroelectrics and magnetic materials dictate the materials response to external field.}
\end{center}
\end{figure}

\paragraph{Defects in solids.}Defects in solids are classified as zero-dimensional (point defects), one-dimensional (line defects) and two-dimensional (surfaces), see Fig.\ref{FigDefects}.
Point defects introduce a change in the scattering power by interfering the coherence between scattered waves by introducing a disturbance to the scattering 
amplitudes (or lengths) and scatterer positions. In diffraction experiments this effect is 
hardly seen if the defect concentration is small. However, once the concentration increases not only is the scattering power changing via change in average $f$ (substitutional disorder) but 
also bond lengths and angles are spatially varying (displacement disorder). Mathematical treatment of displacement disorder correlated with substitutional disorder in 
solid-solutions is given in ref. \cite{Guinier}. 
Phenomenologically the displacement disorder following the substitutional disorder can be described using the concept of microstrain introduced in 
ref. \cite{Stephens} and it is commonly used to model the $hkl$-dependent line broadening in Rietveld refinement: for instance GSAS \cite{GSAS} 
Rietveld refinement software has an option to use profile functions with microstrain broadening formulated in ref. \cite{Stephens}.
 Table \ref{Defects} summarizes common 
disorder types observed in solids and the characteristic signatures in x-ray scattering and diffraction. 

The models given in Table \ref{Defects} may appear rather specific. However, the tabulated signatures are
also found in more general cases. For instance, the signal related to the displacement disorder increases with increasing reflection index 
and vanishes for small values of $\mathbf{s}$. The signatures can be used for distinguishing different structural models. For example, 
both sinusoidal composition modulation and sinusoidal displacement modulation produce satellite reflections at the same $d$-spacing, 
if the propagation vector is the same. The two cases can be identified by their different intensity dependencies.
Also the case of correlated substitutional and displacement disorder can be distinguished from the plain substitutional and 
plain displacement disorder by the characteristic asymmetric intensity of the satellite peaks of each pair. We note that the \emph{pure} 
size-related line broadening gives the same particle size estimate, no matter which reflection is chosen. 
However, if the broadening is not solely due to the particle size it is important to correctly model the disorder. Section \ref{PZT_case}
considers several cases in which non-periodical displacement and substitutional disorder are correlated and the results are 
found to be consistent with the observations seen in periodically modulated cases. 

Model \#1 is a common and straightforward way to describe scattering from a well-mixed solid solution formed by 
not-too-different size constituent elements. The success of the Model \#1 is essentially due to the fact that diffraction tends to emphasize 
average structure and to suppress deviations from it. However, if the composition and atomic scattering factors vary periodically, 
for instance as a sine wave form with a propagation vector $\mathbf{k}$, then the Bragg peaks (principal nodes) are surrounded 
by satellites at distances $\pm \mathbf{k}$ of each Bragg peak. 

In the case of a pure sinusoidal composition variation (in contrast to displacement disorder Models \#3 and \#4 which also exhibit 
characteristic satellite peaks) the ratio of the intensity of the satellite reflections to that of the corresponding Bragg peak intensity 
is a constant for all the nodes. Following ref. \cite{Guinier}, if atomic scattering factor varies as a sine wave, 
$f_n=f(1+\eta\cos 2\pi\mathbf{k}\cdot\mathbf{x}_n)$, then the diffracted wave amplitude is 
$\sum f_n\exp(-2\pi i \mathbf{s}\cdot\mathbf{x}_n)=\sum f \exp(-2\pi i \mathbf{s}\cdot\mathbf{x}_n)+\frac{f\eta}{2}\exp[-2\pi i(\mathbf{s}-\mathbf{k})\cdot\mathbf{x}_n]+\frac{f\eta}{2}\exp[-2\pi i(\mathbf{s}+\mathbf{k})\cdot\mathbf{x}_n]$.
Thus, diffraction pattern exhibits two satellites at distances $\pm \mathbf{k}$ of each node, with intensities which are $\eta^2/4$ times that of the principal node.

Model  \#2 is a well-known description of the atomic displacement originated diffuse scattering implemented in every standard Rietveld refinement software.
Model \#5 is based on the generation of a scattering object from two types of layers $A$ and $B$ with proportions 
$m_A$ and $1-m_A$, respectively. Layers are added to 
the lattice one at a time and the probability of the new layer being an $A$ or $B$ type is dependent only on the preceding 
layer \cite{Welberry}. The scattering object is constructed using conditional probabilities $P(0\vert 0)=1-\alpha$, $P(1\vert 0)=\alpha$, 
$P(0\vert 1)=1-\alpha-\beta$ and $P(1\vert 1)=\alpha+\beta$, where $1$ (0) denotes that the site is occupied by an 
$A$ ($B$)-type layer. In the case of an infinite number of layers rather simple expression is obtained for the diffuse intensity, 
see Model \#5 in Table \ref{Defects}. 

Model \#5 is an example of a model suitable for describing a layer-by-layer crystal growth. 
Contemporary growth chambers are often equipped with the \emph{in situ} monitoring 
facilities, such as Reflection High Energy Electron Diffraction facility allowing to ensure the growth of correct type of layers 
\cite{Biegalski_2008}. Optimization of growth parameters includes the adjustment of substrate temperature, gas atmosphere 
(e.g., oxygen gas pressure) and deposition rate related parameters (such as laser beam fluence in pulsed laser ablation 
deposition or sputtering power). During \emph{in situ} monitoring one observes the diffraction pattern and, if required, conducts 
parameter adjustment until a correct phase is formed. In the case of a single layer it is rather straightforward to observe when 
a correct diffraction pattern emerges. Model \#5 and their extensions are amenable for \emph{in situ} modelling of multilayer thin film structures consisted of stacked 
thin layers possessing different types of layers. 

\begin{sidewaystable}[!p]
\begin{center}
\caption{\label{Defects} Common disorder types and models and characteristic signatures for solids as compiled from 
refs. \cite{Guinier} (\#1-\#4), \cite{Welberry} (\#5),  and \cite{ZevinKimmel} (\#6). Type \#1: The case of a simple 
lattice whose nodes are occupied by atoms $A$, $B$, .... The relative proportions of the atoms are labelled as $c_A$, $c_B$, ... 
and the corresponding scattering factors as $f_A$, $f_B$, ... The average structure factor is $\bar{F}$. The $\mathbf{x}_m$ 
is the vector connecting an atom pair $m$. 
Type \#2: A model for thermal displacement disorder. FWHM, full-width-at-half-maximum, $D$, Debye-Waller factor.
Type \#3: Periodical modulation.
Type \#4: Structure is modulated by a sinusoidal 
distortion, propagation vector $\mathbf{k}$, $\vert \mathbf{k} \vert=\frac{1}{\Lambda}$. Type \#5: One-dimensional 
model for diffuse scattering based on the nearest-neighbour Markov chain. $K$ is a constant determined by $m_A$ and 
the layer form factors $F_A$ and $F_B$. Type \#6: Dislocations.}
\begin{tabular}{l l l l l}
\hline
Model & Disorder type         & Mathematical model                                                                                                                    & Signature                                                                                                                   & Examples                                                          \\
\hline
\#1    & Substitutional          & $\bar{F}=c_Af_A+c_Bf_B+\cdots$                                                                                           & Diffraction pattern is similar to the                                                                           & Solid-solutions of similar size                          \\
          &                                 & $I_2=\Phi_0+$                                                                                                                           & one possessing a translational                                                                                  & atoms. The most common                             \\
          &                                 & $+2\sum_1^\infty+\Phi_m\cos(2\pi \mathbf{s}\cdot \mathbf{x}_m)$                                    &  symmetry: instead of elemental                                                                               & way to model solid-solutions                               \\
          &                                 & $\Phi_0=\overline{f_n^2}-(\bar{F})^2$                                                                                   & scattering factors certain sites                                                                                 & by the Rietveld refinement.                           \\
          &                                 & $\Phi_m=(f_n-\bar{F})(f_{n+m}-\bar{F})$                                                                               & possess an average scatterer $\bar{F}$.                                                                &                                                                        \\            
          &                                 &                                                                                                                                                    &                                                                                                                                   &                                                                        \\
\#2    & Displacement,         & $F_n=f\exp(-2\pi i \mathbf{s}\cdot\Delta\mathbf{x}_n)$                                                        & Part of the diffraction line intensity                                                                           & Isotropic thermal motion of                            \\
          & Debye-Waller $D$  & $D=\exp(\frac{16\pi^2\sin^2\theta}{\lambda^2} \frac{\overline{\Delta x_n^2}}{3})$        &  is shifted to the background. FWHM                                                                         & an atom.                                                         \\
          &                                &                                                                                                                                                     & value of the peak is unaffected.                                                                                &                                                                       \\
\#3    & Displacement          & $\Delta x_n=\frac{\Lambda \varepsilon}{2\pi}\cos(\frac{2\pi na}{\Lambda})$                       & A pair of satellite peaks of equal                                                                                & Nonhomogeneous Cu-Ni-Fe                            \\
          &                                &                                                                                                                                                     &  intensity $\frac{\Lambda^2\varepsilon^2l^2}{4a^2}$ are surrounding               & alloys.                                                            \\
          &                                &                                                                                                                                                     &  each reflection $l$.                                                                                                    &                                                                        \\
\#4    & Correlated             & $f_n=f(1+\eta\sin\frac{2\pi na}{\Lambda})$,                                                                           & Ratio of the intensities of the satellites                                                                     & Solid solutions of different                              \\  
          & Substitutional and & $x_n=na+\Delta x_n=na-\frac{\Lambda\varepsilon}{2\pi}\cos\frac{2\pi na}{\Lambda}$       & to the normal node is asymmetric:                                                                              &  size atoms.                                                    \\  
          & Displacement         &                                                                                                                                                     & $(\frac{\Lambda \varepsilon s + \eta}{2})^2$, $s=\frac{n}{a}-\frac{1}{\Lambda}$  &                                                                         \\  
          &                                &                                                                                                                                                     & $(\frac{\Lambda \varepsilon s - \eta}{2})^2$, $s=\frac{n}{a}+\frac{1}{\Lambda}$. &                                                                        \\
\#5    & Substitutional         & Probability of the new layer                                                                                                        &                                                                                                                                    & Layer-by-layer crystal growth                         \\
          &                               &                                              an $A$ or $B$ depends only on the                                               &$I(\mathbf{s})=K\frac{1-\beta^2}{1+\beta^2-2\beta\cos(2\pi \mathbf{s}\cdot\mathbf{a})}$. & when only short range                \\
       &                               &                                                                                                        immediately preceding layer.  &                                                                                                                                       &   forces are important.                                     \\
\#6    & Dislocation             &                                                                                                                                                      &  At high dislocation densities peak                                                                              &                                                                     \\
          &                                 &                                                                                                                                                      & broadening $\propto \tan\theta$.                                                                             &                                                                        \\       
\hline
\end{tabular}
\end{center}
\end{sidewaystable}

\begin{figure}[h!]
\begin{center}
\includegraphics[angle=0,width=12cm]{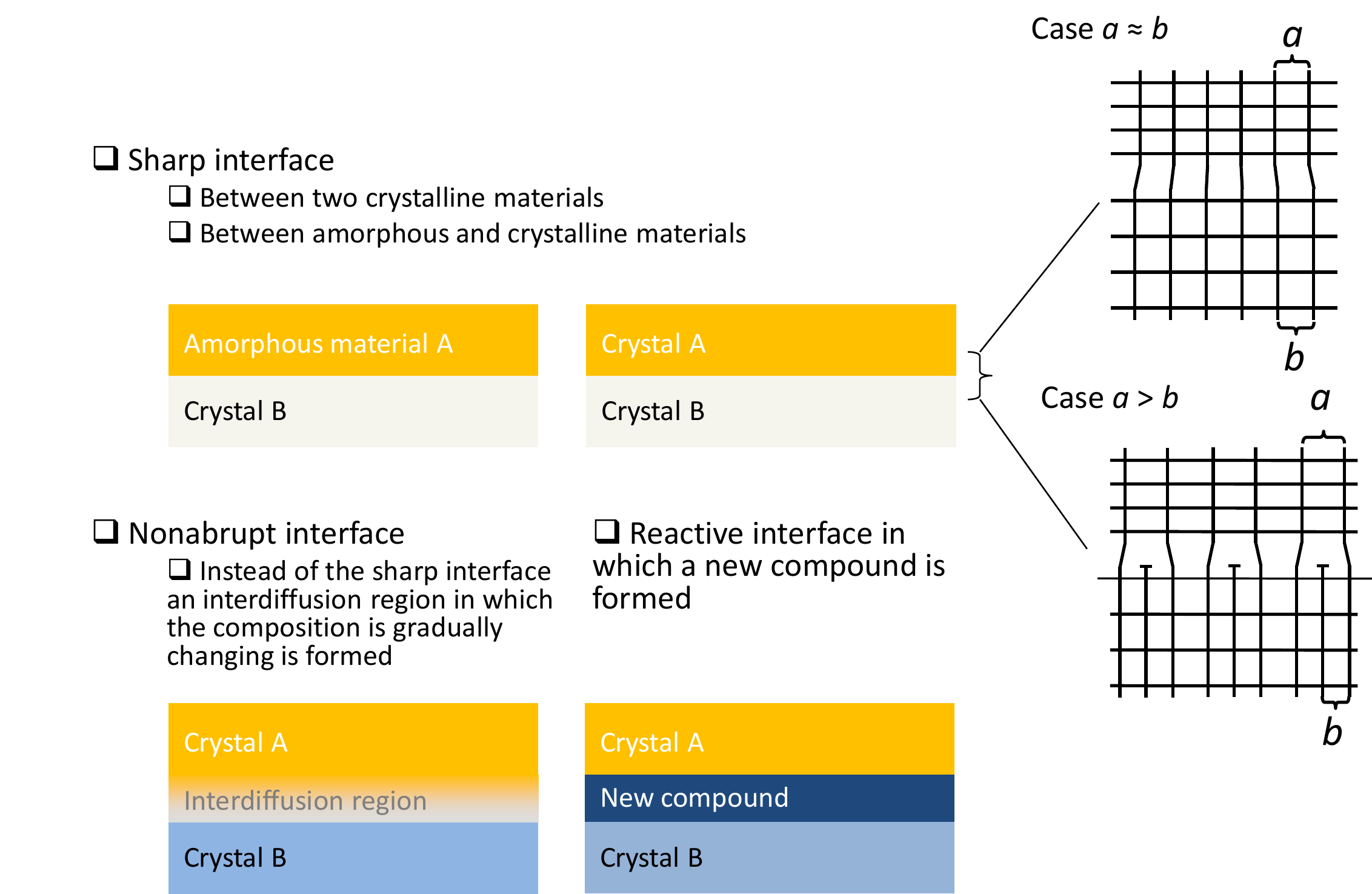}
\caption{\label{ThinFilms}Different interface types formed between material layers.}
\end{center}
\end{figure}

Dislocations come in many forms as do their structural models. Model \#6 summarizes the angle dependent features for high dislocation densities.
Dislocations possess technologically interesting problems as they are often strongly responding to external 
stimulus and thus are time-dependent. The atomic scale structure is not only different in dislocation core but the atomic positions close to a dislocation are 
different from the surrounding matrix \cite{Phillips,KellyKnowles}, making them visible through scattering techniques. In thin films edge dislocations have a 
crucial role as they are favoured over the stressed thin film state once the mismatch between the substrate and thin film exceeds a threshold value 
\cite{Kolasinski,Luth}, see Fig. \ref{ThinFilms}. The plasticity of metals is explained in terms of dislocations. Dislocation cores also serve as a diffusion path in many materials.  

\begin{figure}[h!]
\begin{center}
\includegraphics[angle=0,width=12cm]{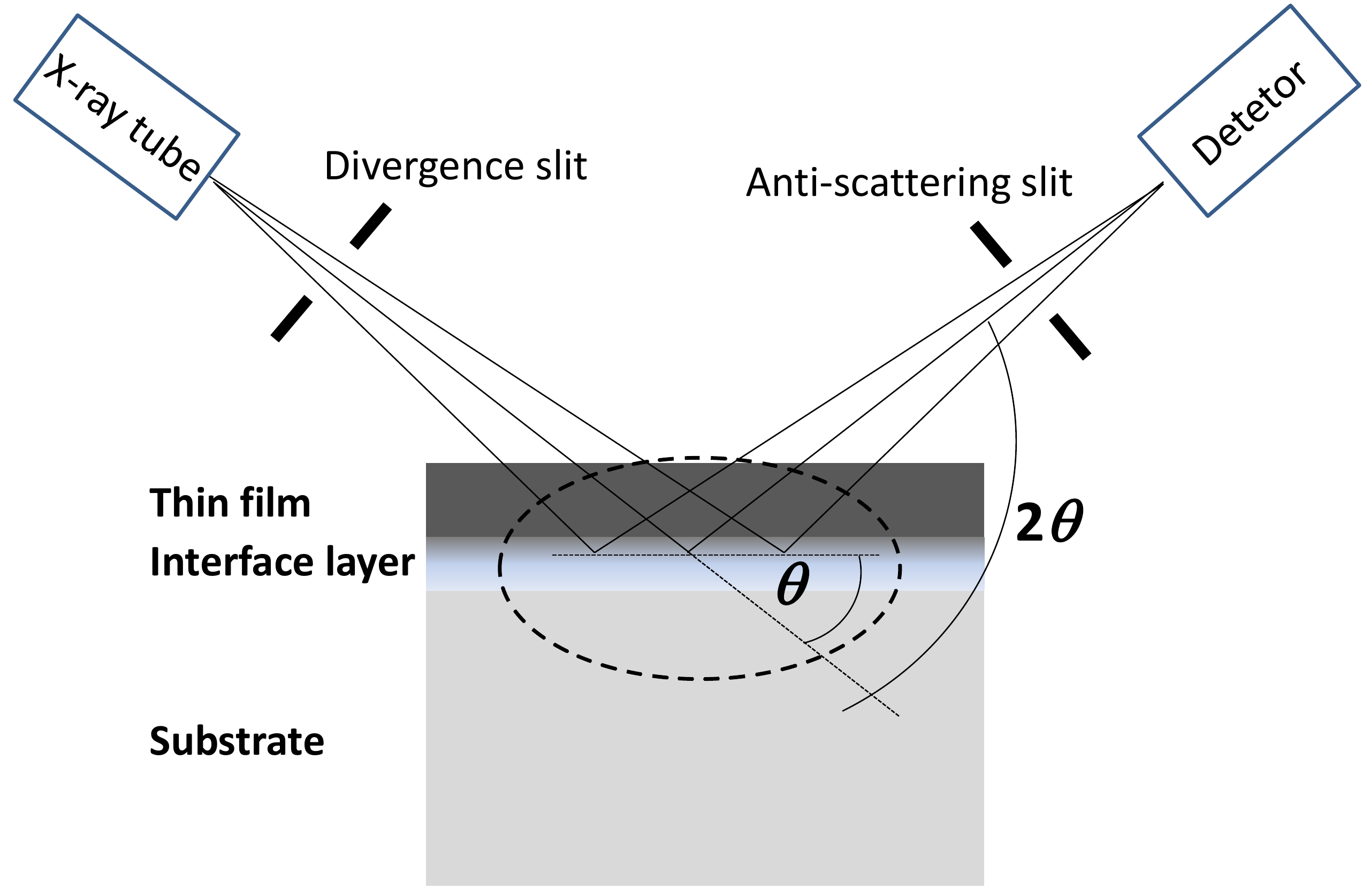}
\caption{\label{XRD_ThinFilm}Schematic picture of a common $\theta-2\theta$ measurement geometry. X-rays penetrate into the substrate 
and the signal is originating from the thin film, interface layer and the substrate. Common approximation is to correspondingly divide the signal 
into three parts, though this is not necessarily justified.}
\end{center}
\end{figure}

Fig. \ref{XRD_ThinFilm} shows schematically why the models applied to data collected on 
commonly used Bragg-Brentano geometry can give 
misleading information in the case of multilayers. First, the net intensity is not a simple superposition of the intensities scattered from individual layers and substrate but involves 
interfacial layers. Second, especially in the case of epitaxial thin films the scattering should be considered to take place in a large entity formed by different layers: in terms 
of the Eq.(\ref{IN}) the summation involves all the atomic pairs in the multilayer structure.
In-situ scattering measurements on crystal growth is becoming a routine experimental techniques and thus models for layer-by-layer growth are 
required. Modelling techniques based on the Markov chain and Ising models are described in ref. \cite{Welberry}.

Similarly the structure in the vicinity of the domain wall separating the different domains also has an atomic scale structure different from either of the domains.
In 180$^{\circ}$ domain boundary the polarization direction is smoothly reversed \cite{Strukov} which is accompanied by a spatial variation of atomic positions. 
The modelling task of multilayer thin films and domain structures is similar and is given in section \ref{DomainSection}.

\subsection{\label{pdf}Pair distribution function method}
Even though isotropic disordered materials, such as a glass or an amorphous material, 
do not have a long-range order, they nearly always possess short-range order. Neutron and x-ray scattering techniques are commonly 
used for determining the distribution function $P(x)$ giving the statistics of atom pairs. Following the treatment given in ref. \cite{Guinier}, 
the principle of the pair distribution function (PDF) 
method can be formulated by expressing the interference function $S(\mathbf{s})$ in terms of the scattered intensity $I_N(\mathbf{s})$ 
and structure factor $F(\mathbf{s})$, forming the average over all orientations of the vector $\mathbf{x}$ connecting the atoms of the pair with 
respect to the scattering vector $\mathbf{s}$ (the sample is assumed to be isotropic), and expressing the interference function 
in terms of the pair distribution functions, Eqs. (\ref{Interference}):
\begin{center}
\begin{equation}\label{Interference}
S(\mathbf{s})=\frac{I_N(\mathbf{s})}{NF^2(\mathbf{s})}=1+\rho_0\int_0^{\infty}4\pi x^2[P(x)-1]\frac{\sin(2\pi sx)}{2\pi sx}dx,
\end{equation}
\end{center}
where $\rho_0$ is the number of atoms of all types per unit volume. 

The reverse Fourier transform of the interference function gives $\rho_0x[P(x)-1]=2\int_0^{\infty}[S(\mathbf{s})-1]\sin(2\pi sx)sds$, which thus would 
require that $S(\mathbf{s})$ is experimentally determined in all of reciprocal space. Instead, a common way is to construct a structural 
model and compute the PDF function, compare it with the experimental data and optimize the model parameters to minimize the difference 
between the computed and measured intensity.
Rather recently, a new method for the calculation of x-ray and neutron powder diffraction patterns from the Debye scattering equation 
was given in ref. \cite{Thomas}. PDF functions were computed as an intermediate stage for computing the diffraction patterns. 
The method is based on the splitting of pairwise atomic interactions into two contributions, the first from lattice-pair vectors and the second from cell-pair vectors.
Illustrative application examples of the PDF method can be found from ref. \cite{Dove}, with an emphasis on the neutron scattering studies of silica.

In order to contrast the method described below and the PDF method we note that the latter suits well for extracting information 
from homogeneously disordered materials, whereas the present method is aimed to provide information about spatially confined 
defects, such as domain boundary. In the present work we also pinpoint the spatial location of the defect.

\subsection{\label{pzt_structures}Solid-solution with correlated substitutional and displacement disorder: lead-zirconate-titanate}
Despite its long-history \cite{Jaffe} PZT is still an intensively studied 
ferroelectric oxide \cite{WebofScience} exhibiting exceptionally high piezoelectric properties when the amount of titanium and zirconium is roughly 
equal \cite{Jaffe,LinesGlass}. Numerous space group assignments for nominally similar samples can be found in the literature (for reviews, see 
refs. \cite{JAC_2014,Frantti_2008}), which is essentially due to the local-scale disorder modelled by different low-symmetry structures. 

\paragraph{Average symmetries.}PZT has a perovskite $AB$O$_3$ crystal structure in which the $B$-cation site is statistically 
occupied by Ti and Zr ions. Metrically, the structure is close to a cube in which Pb cations are approximately at the cube corner, Zr and Ti cations 
are at the cube centre and oxygen anions are close the cube face centres. A well-known summary of the composition and 
temperature dependent phases of PZT is given by the phase diagram of ref. \cite{Jaffe}. Titanium-rich structure is traditionally modelled by 
assuming an ideal space group symmetry ($P4mm$ for $x \leq 0.52$), which is achieved by placing a 
compositionally averaged 'pseudo-atom' at the $B$ site. The anions and cations are displaced from the centrosymmetric positions along positive and negative $c$-axis direction, respectively. 
For Zr-rich concentrations the average structure is rhombohedral, the anion and cation displacements being along the cube diagonal. Notably challenging is the $x\approx 0.50$ composition at which the 
two phases co-exist \cite{Cao}. The two-phase co-existence is typically modelled by refining the total intensity by the superposition of the intensities from the pseudo-tetragonal and pseudo-rhombohedral phases.

The average crystal structures of PZT as a 
function of $x$ can be summarized thus: at room temperature the crystal structure remains tetragonal up to 
$x \approx 0.52$ at which composition (termed the morphotropic phase boundary, MPB) a rather complex set of phases 
emerges, including a monoclinic $Cm$ phase \cite{Noheda,Yokota,Frantti_Neutron} and co-existing low and high-temperature rhombohedral 
phases \cite{Yokota,Frantti_Neutron} . As pointed out in ref. \cite{Glazer_2004} there is no real boundary and the phase co-existence region is extended close 
to the PbZrO$_3$ end of the solid-solution system. It is evident that the $Cm$ phase exists as judged by several high-resolution 
powder diffraction studies, but the phase transition mechanism resulting in an average $Cm$ phase, and even the stability of the
 phase, is still under investigation as it is often linked to the extraordinary piezoelectric properties of PZT at the MPB composition. 
A thermodynamical study showed that an 8th-order Devonshire theory is required to explain the monoclinic phase \cite{Vanderbilt}, 
suggesting that the transition is quite unusual. This was also discussed in ref. \cite{Bell}, where a two-order-parameter 
thermodynamic model was developed for PZT to account for its peculiar features. An early report of the $Cm$ phase in a 
ferroelectric perovskite oxide is about PbNb$_{0.5}$Fe$_{0.5}$O$_3$ (PNF) compound \cite{Lampis,Bonny}. In the 
case of PZT the $Cm$ phase has often been claimed to favour polarization rotation and further to be responsible for the good 
piezoelectric properties. We note that also in the case of PNF the Fe$^{3+}$ and Nb$^{5+}$ cations are disordered. 
The fact that the monoclinic symmetry only tells that the polarization vector is within a mirror plane $m$ does not mean that 
the polarization can rotate within $m$. Nevertheless, it has been proposed that a continuous polarization rotation between 
the $[001]$ and $[111]$ directions along the plane the two direction vectors span would be energetically preferable \cite{Fu}. 
However, we do not adopt that view as detailed in refs. \cite{Frantti_2008} and \cite{Frantti_2008_B}. Instead, to understand 
the polarization reversal we focus on the domain wall and wall motion, 
which involve fairly complex time-dependent structural changes \cite{LinesGlass}.

At higher values of $x$ the dominant phase can, to a good approximation, be described as rhombohedral, there being two variants, with ($R3c$, at low temperatures) and 
without ($R3m$, at high temperatures) octahedral tilts. More precisely, a recent PDF and Rietveld refinement 
study has shown that Zr-rich Pb(Zr$_x$Ti$_{1-x}$)O$_3$ powders possess mixed phases, described by $R3c/R3m+Cm(M_B)$ 
model for  $0.65<x<0.92$ and $R3c/R3m+Cm(M_A)$ model for $0.52<x<0.65$, where $M_A$ and $M_B$ refer to two 
polarization direction variants of the $Cm$ phase \cite{Zhang}.
Neutron diffraction \cite{Phelan} and high-resolution x-ray 
diffraction \cite{Gorfman} studies verified the rhombohedral symmetry and also revealed the presence of the monoclinic phase 
(assigned to $Cm$ symmetry in ref. \cite{Phelan}), providing further support to the idea that the monoclinic phase is not due to 
the presence of adaptive phases. At the highest values of $x$, an antiferroelectric orthorhombic phase is formed. 

\paragraph{Deviations from average symmetries: Signatures of disorder in PZT.}Though the crystal structure models are sufficient for explaining an average structure, they are insufficient when local scale structure is considered. For instance, 
Raman measurements reveal that the spectra are not consistent with the average crystal structure\cite{Frantti_PRB_1999,Frantti_JAP_2013,Frantti_IOP_2014}: the number of Raman active modes is about 
twice the number corresponding to an ideal structure. In diffraction experiments disorder affects the intensity distribution: The fraction of inelastically scattered intensity is increased (as discussed 
 below in the context of atomic displacement parameter (ADP)) and the tail regions of the Bragg reflection gain intensity. 
The displacement disorder can be divided into Pb displacements (off-site $A$-cations) and different 
positions of the $B$-cations, Zr and Ti \cite{Glazer_2004,Corker,Ricote, Muller,Frantti_2002,Goossens}. Large off-centre 
Pb displacements are common in perovskite oxides \cite{Lampis,deMathant,Malibert,Dkhil,Pasciak}. 
Randomly distributed Zr and Ti atoms cause locally varying bond lengths and 
angles, frequently approximated by introducing so called microstrain. In contrast to the ADP induced diffuse scattering the reflection widths are $hkl$-dependent in microstrained samples. A random distribution of Zr and Ti results in corresponding 
distribution of Pb displacements. 
If the point defect concentration is sufficiently small (of the order of $x =0.01$ or less) so that the long-range order characteristic to the crystal 
symmetry is not changed, defects mainly contribute to the intensity of the tail areas around the Bragg reflections: diffuse 
scattering is increased, whereas elastic scattering is decreased. The effect is most clearly seen at small $d$-spacing area 
(see, e.g., ref. \cite{ZevinKimmel}). Correspondingly, point defects are frequently seen as abnormal ADP's, which are either 
anomalously large or unphysically small: even diagonal components can be negative. This 
issue is addressed in ref. \cite{Massa}. The negative values can be due to the fact that ADP's try to model two separate types of disorder,
 dynamic (thermal vibrations) and static (substitutional and/or positional disorder). Small substitute atomic concentrations 
are seldom capable of introducing large changes in x-ray scattering intensities (unless there is a large difference in 
the numbers of electrons of the atoms) though they may result in detectable changes in neutron scattering. Point defect concentrations of a fraction 
of  at.\% may, however, result in observable changes in bond lengths \cite{Frantti_PRB_1997}. 

\begin{figure}[h!]
\begin{center}
\includegraphics[angle=0,width=12cm]{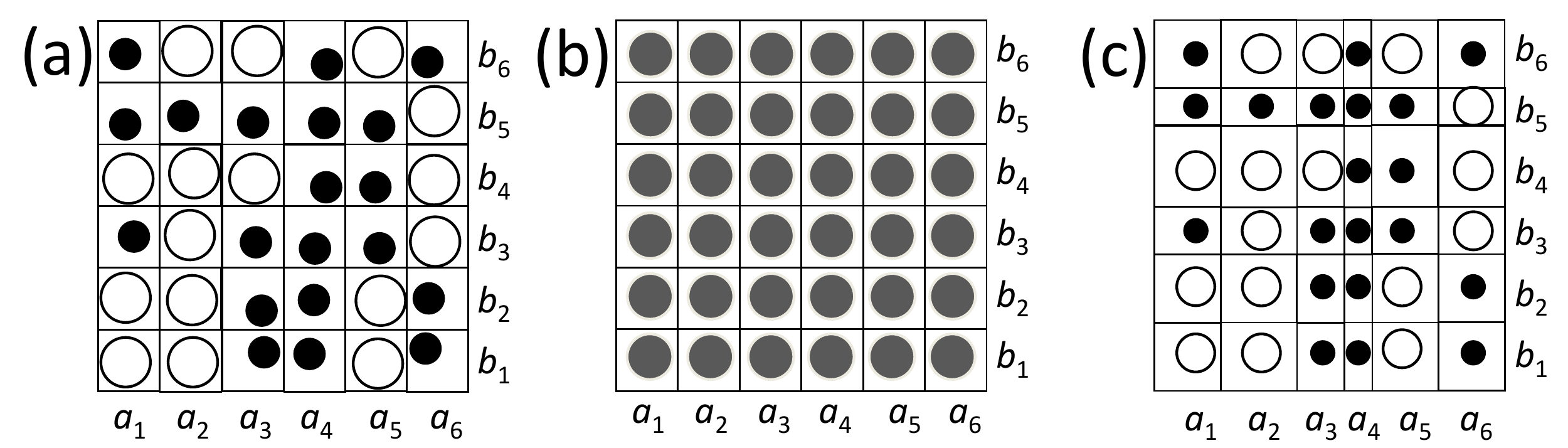}
\caption{\label{Solidsolution} (a) Schematic illustration of a solid-solution, (b) typical averaging and (c) an approximation which takes the local disorder into account.}
\end{center}
\end{figure}

At large point defect concentration it is better to abandon the concept of well-ordered host crystal with point defects and to 
model the system from the beginning. Fig. \ref{Solidsolution} schematically illustrates a binary solid solution and the commonly used 
approximation in which one does not make a distinction between the two types of atoms, but only considers a pseudoatom 
taken to possess a scattering amplitude formed as a composition weighted average of the scattering amplitudes of the two atoms. 
Three-dimensional  periodicity is commonly introduced in a similar manner. Thus, the disorder is averaged away as it is a heavy task 
to compute the scattering intensity corresponding to a huge unit cell. Below we introduce an approximation which 
keeps the essential features of the solid-solution, namely the inhomogeneous distribution of two types of atoms and the 
corresponding variation in bond lengths, corresponding to the case illustrated in Fig. \ref{Solidsolution}(c).

\section{\label{PZT_case}Scattering from PZT clusters}
A method for modelling the local structure under the constraint that the 
\emph{average} structure remains intact was recently developed \cite{Patent} and applied for addressing the 
cation displacements as a function of hydrostatic pressure \cite{Frantti_2014}. Central part of the modelling work is the parameterization of the disorder. 
As in the case of crystals with well-defined space group symmetries there is a requirement to fill the space exactly once. 
This sets limits to the cells and their mutual connectivity. A straightforward way to do this is given below.
Focus is on the 
construction of a model which takes the disorder into account yet is computationally sufficiently simple. 
We first describe how a single scattering cluster is constructed. 
The material to be modelled can entirely be consisted of a single scattering cluster, or as a special case the cluster can 
be a unit cell in a crystallographical sense. 

Next section introduces the parameters used in this paper after which application examples are given.

\subsection{\label{cluster_section}Generation of the scattering cluster}
Below a construction of a model of a combined substitutional and displacement disorder is given. 
The influence of the crystal size and shape on the peak profiles is taken into account. 
Scattering intensity is computed for single clusters.
The program was written using the C-language and a message passing interface (MPI). 
Computational platform was provided by the 
CSC (Finnish IT Center for Science Ltd.,  administered by the Ministry of Education, Science and Culture).

We first generate a scattering cluster in which each ion is placed on a specific site (described below). 
Ti and Zr are statistically distributed after which the atomic positions are adjusted. 
Specifically, each cell in the cluster is either tetragonal or rhombohedral and the ion positions 
are relaxed accordingly. The relaxed structural parameters are tabulated in Table \ref{SitesandAtoms}. Atomic positions are adjusted by bond-valence-sum method \cite{Brown}. 

\paragraph{Initial parameters.}Experimental room-temperature structural values of PbTiO$_3$ and PbZrO$_3$ are used for generating the initial values for the scattering cluster.
In this stage, also alternative methods could be used.
The lattice parameters for PbTiO$_3$ are: $a_T=b_T=3.9000$ \AA{ }and $c_T=4.1500$ \AA{ }and for PbZrO$_3$ the lattice parameters are $a_Z=b_Z=c_Z=4.138$ \AA. Though 
PbZrO$_3$ is orthorhombic, we average the structure to be cubic (space group $Pm\bar{3}m$). For the computation of the scattering power the positions of the ions are referred to 
points $i,j,k$ which give the origins of the cells. In the special case of a scattering volume possessing translational symmetry the points $i,j,k$ form a crystallographical lattice. Here the 
focus is on the cases lacking translational symmetry. 

The structure of the cells depends on the Zr concentration: at high titanium concentrations the cells are tetragonal, at $x\approx 0.50$ two phases co-exist and at high 
Zr-concentrations rhombohedral cells are dominant. This feature is embedded in the model as described in flow chart \ref{FlowChart}.

Table \ref{SitesandAtoms} lists the sites in tetragonal and rhombohedral  cells and Table \ref{ScatteringParameters} lists the neutron scattering lengths and coefficients used for the computation of x-ray scattering factors.

\begin{table}[htb!]
\begin{center}
\caption{\label{SitesandAtoms} Fractional coordinates of the  atoms in pseudo-tetragonal and pseudo-rhombohedral structures. 
Structural parameters $x$(Pb), $y$(Pb), $z$(Pb), $z$(Zr), $z$(Ti) and $z$(O) were computed for each cell $i,j,k$ by 
bond-valence sum method.}
\begin{tabular}{l l l l}
Tetragonal structure
                                                        &                          &                               &                                     \\
\hline
Atom                                               &   $x$                 &  $y$                       & $z$                              \\
Pb                                                   &   $x$(Pb)           &  $y$(Pb)$=x$(Pb) & $z$(Pb)$=x$(Pb)           \\
Zr                                                    &   $\frac{1}{2}$  & $\frac{1}{2}$      & $\frac{1}{2}+z$(Zr)   \\
Ti                                                    &   $\frac{1}{2}$   & $\frac{1}{2}$      & $\frac{1}{2}+z$(Ti)    \\
$0$I                                                &  $\frac{1}{2}$   & $\frac{1}{2}$      &  $z$(O)                        \\
$0$II                                               &  $0$                   & $\frac{1}{2}$      &  $\frac{1}{2}+z$(O)   \\
$0$III                                              &  $\frac{1}{2}$  & $0$                       &  $\frac{1}{2}+z$(O)   \\
\hline
Rhombohedral structure
                                                        &                                      &                                      &                                   \\
\hline
Atom                                               &   $x$                             &  $y$                              & $z$                              \\
Pb                                                   &   $x$(Pb)                       &  $y$(Pb)$=x$(Pb)       & $z$(Pb) $=x$(Pb)       \\
Zr                                                    &   $\frac{1}{2}+z$(Zr)  & $\frac{1}{2}+z$(Zr)     & $\frac{1}{2}+z$(Zr)   \\
Ti                                                    &   $\frac{1}{2}+z$(Ti)   & $\frac{1}{2}+z$(Ti)      & $\frac{1}{2}+z$(Ti)    \\
$0$I                                                &  $\frac{1}{2}$             & $\frac{1}{2}$               &  $z$(O)                        \\
$0$II                                               &  $0$                             & $\frac{1}{2}$               &  $\frac{1}{2}+z$(O)   \\
$0$III                                              &  $\frac{1}{2}$            & $0$                                &  $\frac{1}{2}+z$(O)    \\
\hline
\end{tabular}
\end{center}
\end{table}
 
\begin{table}[htb!]
\begin{center}
\caption{\label{ScatteringParameters} Coefficients for analytical approximation to the x-ray scattering factors $f$ and 
the neutron scattering lengths $b_0$, taken from ref. \cite{ITCC}. Scattering factors are computed using the approximation 
$f=z+\sum_i a_i\exp(-b_is^2)$, where $s=\sin\theta/\lambda$. Anomalous dispersion coefficients are $f_1$ and $f_2$ 
(real and imaginary parts, respectively), the given values correspond to the Cu K$_{\alpha}$-radiation 
($\lambda = 1.540562$ \AA). The Bragg diffraction angle is labelled as $\theta$ and $\lambda$ is the x-ray wavelength.}
\begin{tabular}{l l l l l}
\hline
Parameter    & Pb       & Zr       & Ti       & O        \\
$b_0$        & 9.405    & 7.16     & -3.370   & 5.803    \\
$z$          & 13.4118  & 2.06929  & 1.28070  & 0.250800 \\
$a_1$        & 31.0617  & 17.8765  & 9.75950  & 3.04850  \\
$b_1$        & 0.690200 & 1.27618  & 7.85080  & 13.2771  \\
$a_2$        & 13.0637  & 10.9480  & 7.35580  & 2.28680  \\
$b_2$        & 2.35760  & 11.9160  & 0.500000 & 5.70110  \\
$a_3$        & 18.4420  & 5.41732  & 1.69910  & 1.54630  \\
$b_3$        & 8.61800  & 0.117622 & 35.6338  & 0.323900 \\
$a_4$        & 5.96960  & 3.65721  & 1.90210  & 0.867000  \\
$b_4$        & 47.2579  & 87.6627  & 116.105  & 32.9089    \\
$f_1$        & -4.075   & -0.186   & 0.219    & 0.049              \\
$f_2$        & 8.506    & 2.245    & 1.807    & 0.032               \\
\hline
\end{tabular}
\end{center}
\end{table}

\paragraph{Flow chart.}Flowchart describing the steps involved in the scattering cluster construction.
\begin{enumerate}
\label{FlowChart}
\item The size of the scattering cluster is given in terms of parallel epipeds, or cells. In the present example 
a rectangular parallel epiped cluster, consisted of $N_x\times N_y \times N_z$ cells, is constructed.  
Each cell is referred to by indices $i=0,...,N_x$, $j=0,...,N_y$ and $k=0,...,N_z$. 
\item Atoms are placed into the initial sites according to Table \ref{SitesandAtoms}. Either Ti or Zr is inserted into 
each cell with probabilities $1-x$ and $x$, respectively. A random number generator is used for this purpose. 
\item The disorder to be modelled is correlated substitutional and displacement type caused by the random distribution 
of Ti and Zr ions. For modelling purpose we use three sum functions $c_x(i)$, $c_y(j)$ and $c_z(k)$. 
$c_x(i)$ gives the number of Zr atoms in planes $i = constant$, $constant$ being  $1,2,...,N_x$. Functions $c_y(j)$ and 
$c_z(k)$ posses similar meaning.
\item \label{CellEdges}Cell edges are readjusted so that the compatibility condition is fulfilled. Fig. \ref{Compatibility} illustrates the 
compatibility in two-dimensional case. 
\begin{figure}[h!]
\begin{center}
\includegraphics[angle=0,width=10cm]{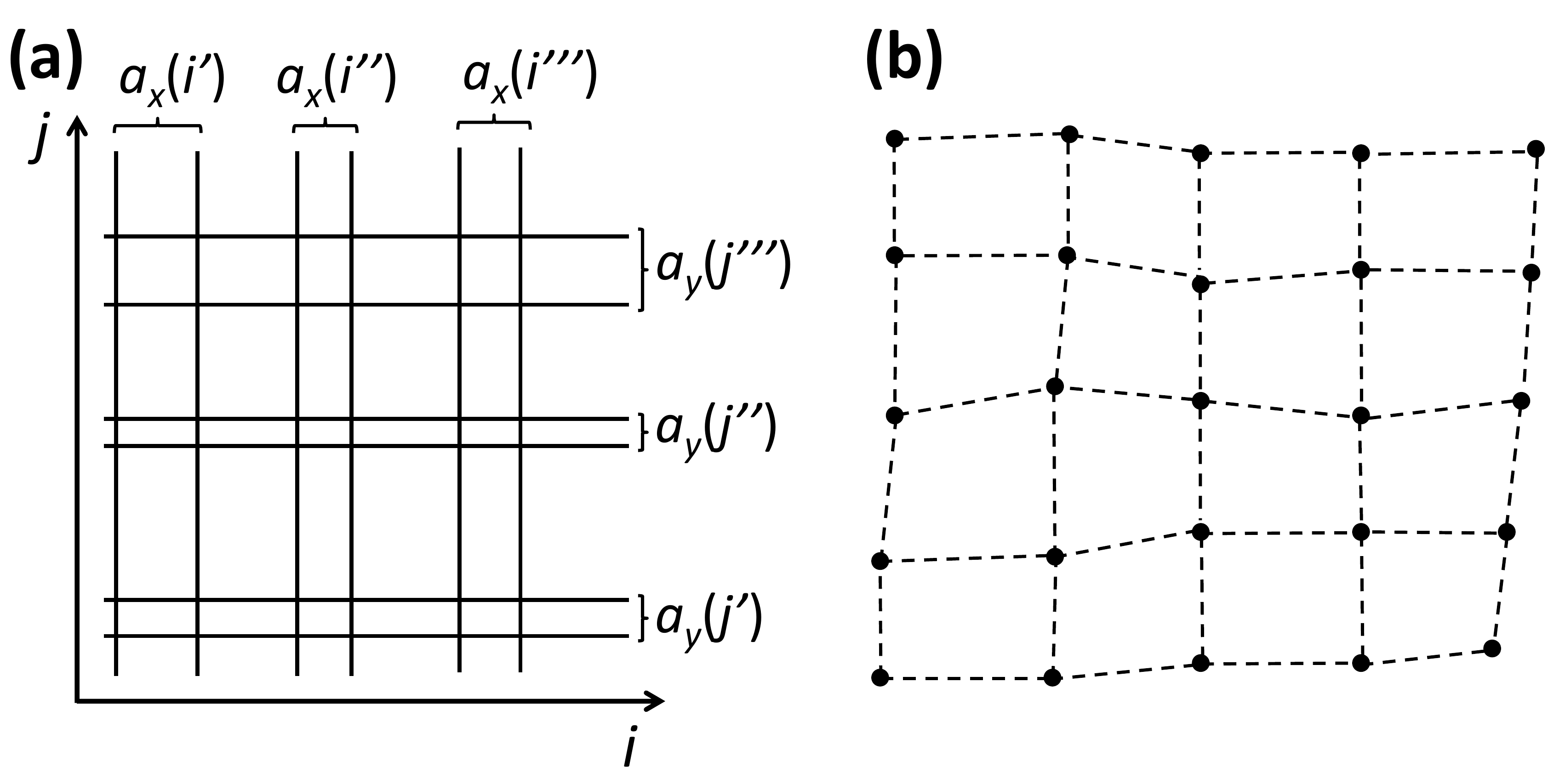}
\caption{\label{Compatibility}(a) The compatibility between adjacent cells is fulfilled with spatially varying cell edge lengths. 
Though the cell edge lengths are spatially varying they are continuously connected and discontinuities, illustrated in (b), are 
avoided. 
For simplicity, the angles are taken to be orthogonal, which however is not a requirement of the method.}
\end{center}
\end{figure}
The cell edges $a_x(i)$, $a_y(j)$, and $a_z(k)$ 
are relaxed either to be rhombohedral or tetragonal:\\
if $c_x(i)\geq N_yN_z/2.0$ and $c_y(j)\geq N_xN_z/2.0$ and $c_z(k)\geq N_xN_y/2.0$ \\
$a_x(i)=(a_T^2c_T)^{1/3}+K_x(c_x(i)+c_y(j)+c_z(k))/3.0*(a_Z-(a_T^2c_T)^{1/3})/(N_yN_z)$\\
$a_y(j)=(a_T^2c_T)^{1/3}+K_x(c_x(i)+c_y(j)+c_z(k))/3.0*(a_Z-(a_T^2c_T)^{1/3})/(N_xN_z)$\\
$a_z(k)=(a_T^2c_T)^{1/3}+K_x(c_x(i)+c_y(j)+c_z(k))/3.0*(a_Z-(a_T^2c_T)^{1/3})/(N_xN_y)$\\
else \\
$a_x(i)=a_T+K_x(a_Z-a_T)c_x(i)/(N_yN_z)$\\
$a_y(j)=b_T+K_x(b_Z-b_T)c_y(j)/(N_zN_x)$\\
$a_z(k)=c_T+K_x(c_Z-c_T)c_z(k)/(N_xN_y)$.\\
Thus, both rhombohedral and tetragonal cells can co-exists within a same cluster. To further simplify the model, the largest rhombohedral cell is determined after which  
all rhombohedral cells are constrained to have this size. This also ensures that all cells are continuously connected to adjacent cells. 
A parameter $K_x$ $(\approx 1)$ is introduced. This is related to the fact that average $B$ valence exceeds 4 in Zr-rich PZT $(x \geq 0.54)$.
To obtain nominal valence the cell volume is gradually expanded by increasing the value of $K_x$ till atoms can be positioned so that they possess nominal valences.

For simplicity, linear relationship between lattice constants and sum functions is assumed. We also assume that 
the same composition variation causes the same structural variation in the $x$ and $y$ directions. This also implies that for a single phase cluster the average values of 
$a_x$ and $a_y$, respectively denoted as $\langle a_x\rangle$ and $\langle a_y\rangle$, are equal. This is because 
$\sum_{i=1}^{i=N}c_x(i)=\sum_{j=1}^{j=N}c_y(j)=N_{\mathrm{Zr}}$, where $N_ {\mathrm{Zr}}$ is the number of Zr atoms in 
the cluster. The relationships are no longer true for clusters possessing co-existing tetragonal and rhombohedral cells. By giving up these 
conditions one could approach more complex cases. We also assume that $x$, $y$ and $z$ directions are orthogonal.
\item Cation positions are adjusted to fulfil the nominal bond-valence sums, described below. 
\item Parallel epiped shaped particles are hardly seen in real materials, so an ellipsoidal cut of the cluster is formed.
The semiaxes of ellipsoids, including spheres, are chosen to be parallel to the cell edge directions.
Rounded cluster has a feature that the subsidiary maxima are diminished, in contrast to the parallel epiped clusters.
\item The scattering power of the cluster is computed by Eq. (\ref{IN}) for selected directions. Directions are given in terms 
of $\mathbf{s}=h\mathbf{a}^*+k\mathbf{b}^*+l\mathbf{c}^*$, where $\mathbf{a}^*=\hat{\mathbf{x}}/\langle a_x\rangle$, 
$\mathbf{b}^*=\hat{\mathbf{y}}/\langle a_y\rangle$ and $\mathbf{c}^*=\hat{\mathbf{z}}/\langle a_z\rangle$, where $\hat{\mathbf{x}}$, $\hat{\mathbf{y}}$, and $\hat{\mathbf{z}}$ are unit vectors parallel to the positive $x$, $y$ and $z$-axis directions, respectively. The scattering power is multiplied by 
Lorentz factor (Eq.(\ref{LN})) in the case of neutrons, and by the $Lp$ factor (Eq.(\ref{Lp})) in the case of x-rays, see Appendix I.
\end{enumerate}

\paragraph{Bond-valence-sum based atomic position adjustment.}
Bond-valence-sum (BVS) method \cite{Brown} is applied to calculate the cation positions with respect to the oxygen polyhedra. Alternative computational 
techniques, such as using empirical potentials for structure optimization, or experimental techniques, could also be used. 
In the model Zr and Ti cations are displaced 
along the $c$-axis direction if the cell is tetragonal, otherwise the 
displacement is along the cell diagonal (the rhombohedral cells). Lead cations, 
consistently with the experimental observations, are always displaced along the cell 
diagonal. 

As the cells are too tight for Zr (there is no position at which the Zr valence 
would be +4), the Zr positions are first adjusted so that Zr valences are minimized, after which the Ti positions are 
adjusted so that the average valence is 4. If the average valence is above 4, 
the cell size is increased by increasing the parameter $K_x$, see item \ref{CellEdges} in flow chart \ref{FlowChart}.

\subsection{\label{spherical_section}Single domain spherical clusters}
\begin{figure}[h!]
\begin{center}
\includegraphics[angle=0,width=12cm]{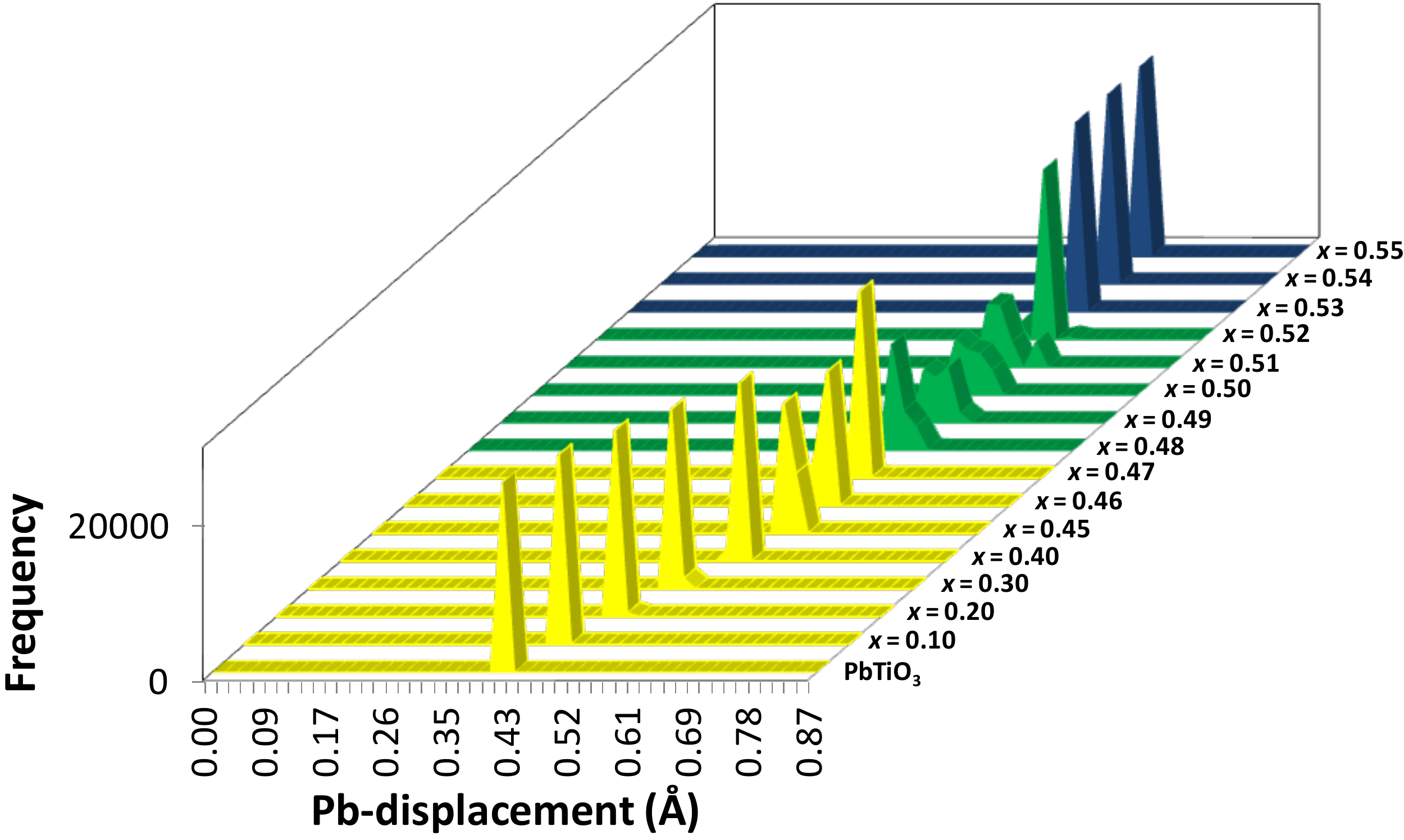}
\caption{\label{A_Displacements} Statistical distribution of the Pb-cation displacements as a function of $x$. Yellow, green and blue colour 
indicates the single phase tetragonal, two-phase and rhombohedral areas, respectively. Worth to note is that the peaks are broadest in the two-phase region.
In each cell Pb-cations are displaced along the cell diagonal. Pb-displacements increase with increasing $x$.}
\end{center}
\end{figure}

\begin{sidewaysfigure}[!p]
\begin{center}
\includegraphics[angle=0,width=20cm]{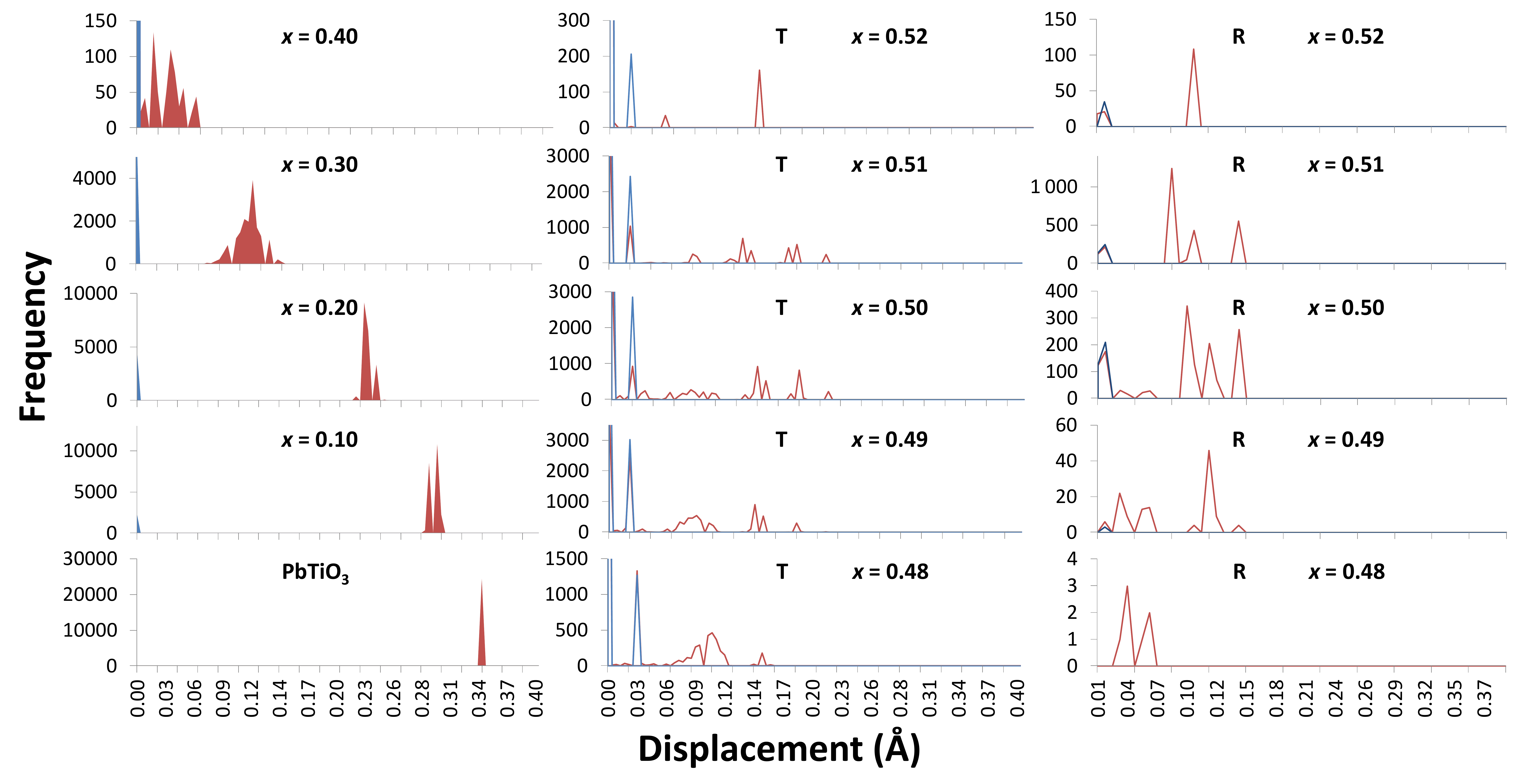}
\caption{\label{B_Displacements} Statistical distribution of the Zr and Ti-cation displacements from oxygen octahedra centre as a function of $x$. 
Red and blue colour indicates Ti- and Zr cations positions, respectively. The left-hand column gives the displacements in the single phase region, 
whereas the centre and right-hand columns give the displacements in the two-phase co-existence region. Left-hand column shows how the cation 
distribution broadens with increasing $x$, in strong contrast with  models based on the translational symmetry. Worth to note is that not only Zr but 
also Ti displacements are rather small at Zr-rich compositions (note the scale differences in Frequency axis). This suggests that the Pb-displacements 
significantly contribute to the dipole moment at MPB. The role of Zr is to stretch the cells so that Pb ions are forced to be significantly 
displaced from cuboctahedra centres in order to achieve nominal valence. See also text.}
\end{center}
\end{sidewaysfigure}

To address the local structural changes and microstrain spherical clusters with a radius of 18 cell lengths ($N_x=N_y=N_z=36$) are generated.
\paragraph{Correlated substitutional and displacement disorder.}PZT is evidently a material in which displacement disorder follows substitutional disorder: volumes 
with high Zr-concentration have larger cells. The distribution of $B$-cations itself is assumed to be random as described in flow chart \ref{FlowChart}.
Figures \ref{A_Displacements} and 
\ref{B_Displacements} show the $A$- and $B$-cation displacements as a function 
of $x$. 
Model indicates rather small displacement of Zr and Ti cations from the oxygen octahedra centre at and in the vicinity of the 
MPB region. The model predicts that the ferroelectric polarization is essentially due to the Pb and Ti cation displacements. Above 
$x>0.40$ it is essentially the Pb-displacements which are responsible for polarization in PZT. The increasing width of the 
cation displacement distribution with increasing $x$ indicates larger deviation from the average symmetry: In the case of perfect translational 
symmetry all $A$- and all $B$-cation displacements would be peaked at a single value. Also the number of displaced $B$-cations 
is strongly decreasing with increasing $x$: The $B$-cation displacements are centred close to the oxygen octahedron centre when $x \approx 0.40$. As 
Fig. \ref{A_Displacements} shows, the width of the Pb-cation displacements is largest in the two-phase region (green data). It is also evident that 
the Pb-ion displacements are increasing with increasing $x$.
As was discussed in ref. \cite{Frantti_2014}, the $B$-cation displacements are probably underestimated: If one gives up the constraint that 
the average valence of $B$-cations should be $+4$ and that the Pb-cation should have a valence of $+2$ and replaces it with a less severe constraint 
that the sum of the two types of cations should be $+6$ then the $B$-cations can be displaced by a larger amount also at Zr-rich areas. This would be 
compensated by a smaller Pb-displacements. However, such a computation would require an energy minimization. The BVS values do not necessarily 
correspond to the energy minimum, though it is reasonable to assume that they are not too far off as the BVS parameters are based on 
a fit to a vast number of experimental data. It is reasonable to assume that most reported structures correspond to the energy minimum.

\begin{sidewaysfigure}[!p]
\begin{center}
\includegraphics[angle=0,width=20cm]{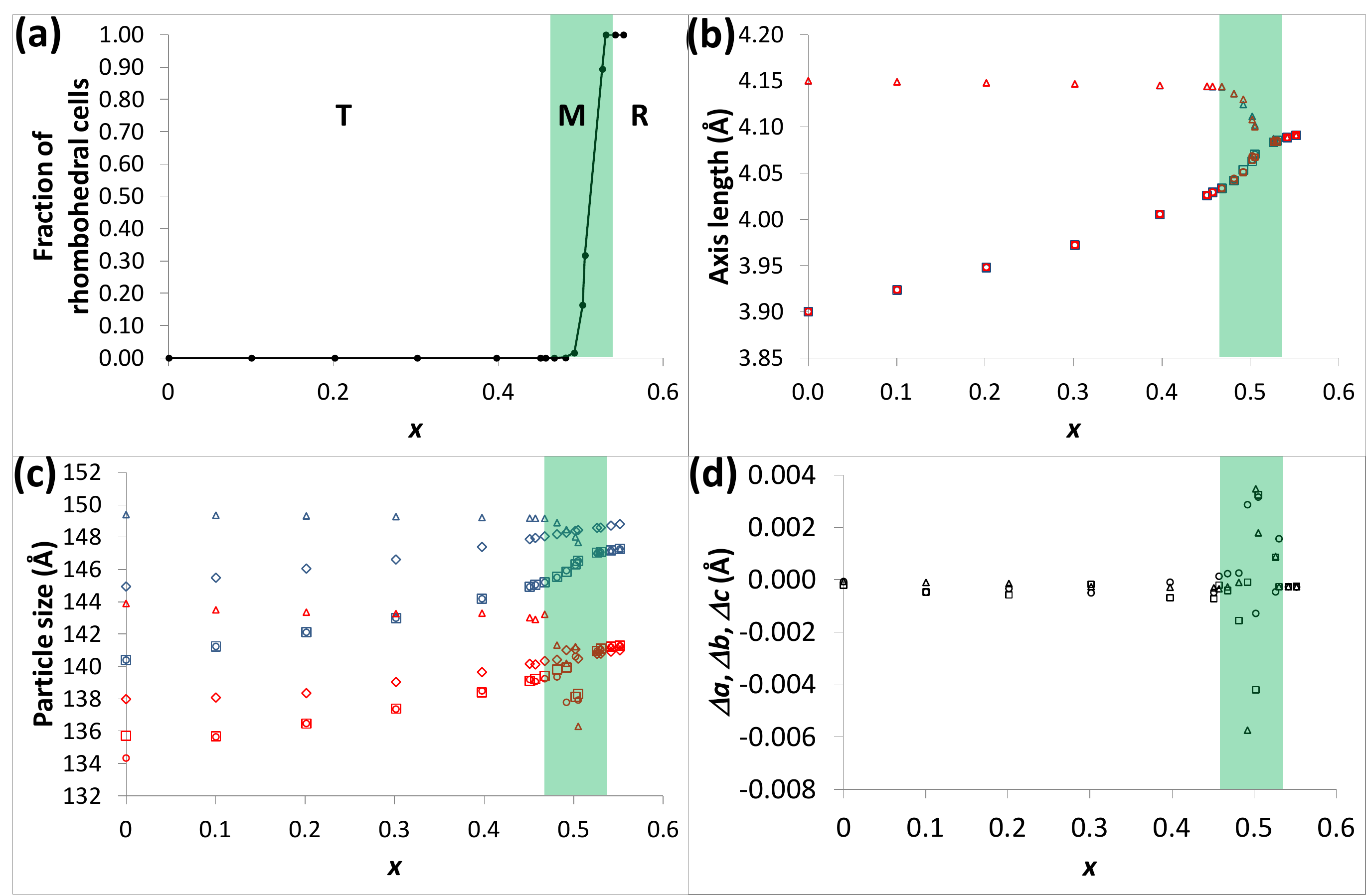}
\caption{\label{Statistics}Composition dependent fraction of the rhombohedral cells, panel (a), average lattice parameter values taken from the known cluster parameters (blue data points) 
and average lattice parameter values obtained from the Bragg equation (red data points), panel (b), cluster dimensions 
(cluster values are given by blue markers and values obtained from the Scherrer equation are plotted by red markers), panel (c) and the difference between the true 
and lattice parameters obtained through the Bragg equation, panel (d). Squares, circles and triangles indicate the $a$, $b$ and $c$ axes values. Red and blue diamonds (panel (c)) 
correspond to the values extracted from the $222$ reflection positions and the sphere dimension along the $\langle111\rangle$ direction, respectively. 
Green shadow indicates the two-phase co-existence region, also referred to as the MPB region. Letters T, M and R stand for the tetragonal, mixed and rhombohedral region, respectively. }
\end{center}
\end{sidewaysfigure}

\begin{figure}[!p]
\begin{center}
\includegraphics[angle=0,width=13cm]{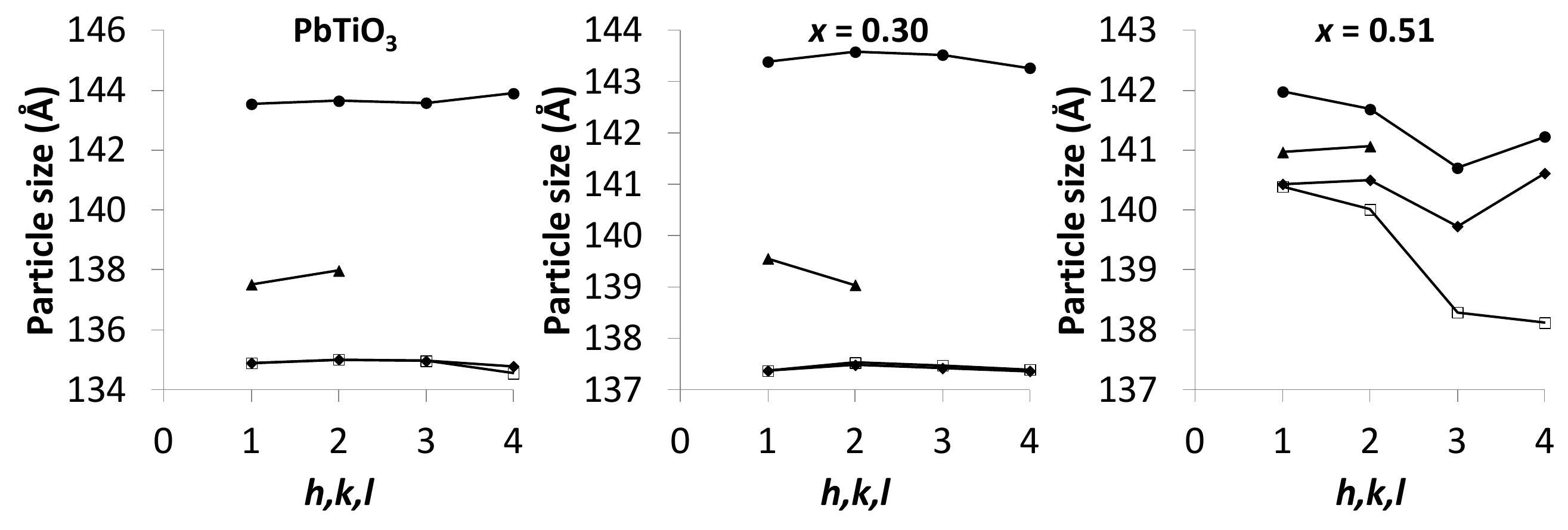}
\caption{\label{Microstrain}Particle size as estimated from the apparent full-width-at-half-maximum values of the x-ray scattering reflections. Open squares, filled diamonds, filled circles and filled triangles
are values estimated from the $h00$, $0k0$, $00l$ and $hhh$ reflections, respectively. Only $x=0.51$ cluster shows evident reflection indice dependent broadening, implying that the broadening 
is not solely due to the particle size.}
\end{center}
\end{figure}

\paragraph{Cluster and cell dimensions.}Fig. \ref{Statistics}(a) shows the number of rhombohedral cells 
as a function of $x$. The model is seen to be consistent with the known MPB composition (green 
shadows in Fig. \ref{Statistics}(a)-(d)), as the strong increase onset at $\approx 0.47$ shows. This implies that relatively 
straightforward statistical approach (given in flow chart \ref{FlowChart}) is capable of explaining the two-phase co-existence. The two-phase
coexistence within a same cluster affects the line-shapes in the two-phase regions, resulting in less accurate lattice parameter 
values in the MPB region, see Fig. \ref{Statistics}(d). 
Fig. \ref{Statistics}(c) plots the cluster dimensions and compares the dimensions to the values obtained through the Scherrer equation:
\begin{equation}\label{Scherrer}
D_p=4/3(\pi/6)^{1/3}\lambda/(\theta_{fwhm}\cos\theta)
\end{equation}
where $\theta$ and $\lambda$ are the Bragg peak centre position and wavelength values and $\theta_{fwhm}$ is the full-width-at-half-maximum 
of the Bragg peak. All angles are given in radians. The constant is characteristic to the spherical shaped particles.
The maximum dimension of the cluster is slightly larger than the value obtained from the Scherrer equation. This is partially due to the fact that the 
cluster is not exactly a sphere but is consisted of pseudocubic cells so that the maximum dimension, given by blue markers in Fig.  \ref{Statistics}(c)
are slightly larger than the diameter of the spherical surface fit to go as close to the cluster exterior as possible. We note that the line broadening 
due to the particle size effect alone yields an apparent size which is independent of the order of the reflection, while it depends on the reflection order in 
correlated substitutional and displacement disorder. This is also illustrated in Fig. \ref{Microstrain} which plots the particle size, estimated through Eq. (\ref{Scherrer}) for PbTiO$_3$ and $x=0.30$ 
and $x=0.51$ PZT compositions from 14 reflections. The effect of combined substitutional and displacement disorder is most evident in the $x=0.51$ sample,
seen as a large number of satellite peaks (though there is no simple modulation of a periodic structure, we still use the term satellite to distinguish the subsidiary 
minima peaks due to the limited cluster size and disorder generated peaks).

\begin{figure}[!p]
\begin{center}
\includegraphics[angle=0,width=13cm]{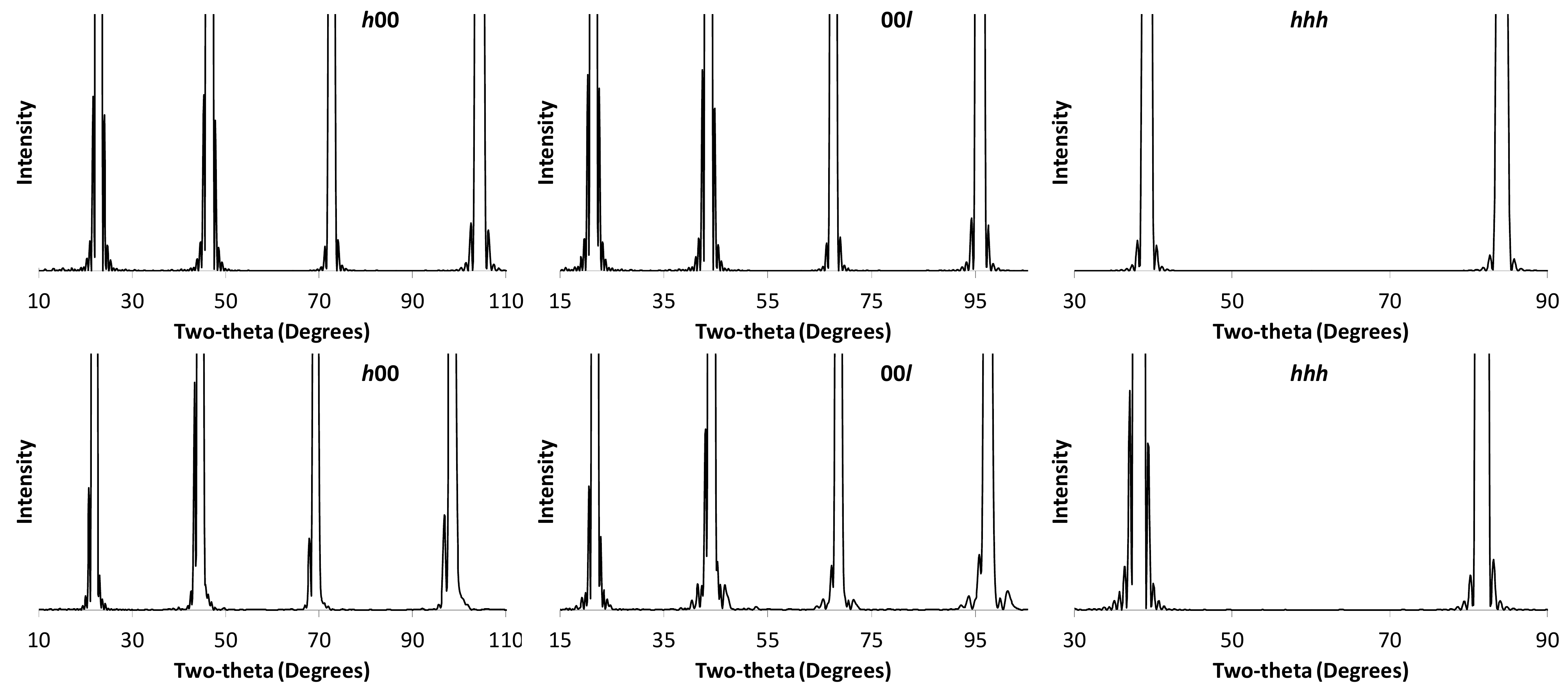}
\caption{\label{Satellites}Neutron scattering intensity of the $h00$, $00l$, and $hhh$ type reflections. Upper row shows the intensity computed for the PbTiO$_3$ cluster and the lower row 
the intensity computed for the $x=0.51$ cluster. The combined substitutional and displacement disorder is evident in $x = 0.51$ cluster, as is seen from the numerous satellite 
reflections. Also notable is the asymmetry of the intensity of the satellite peaks on the larger and smaller $2\theta$ side of the main peak. The asymmetry is much larger than the 
impact of the Lorentz factor.}
\end{center}
\end{figure}

\begin{figure}[!htp]
\begin{center}
\includegraphics[angle=0,width=13cm]{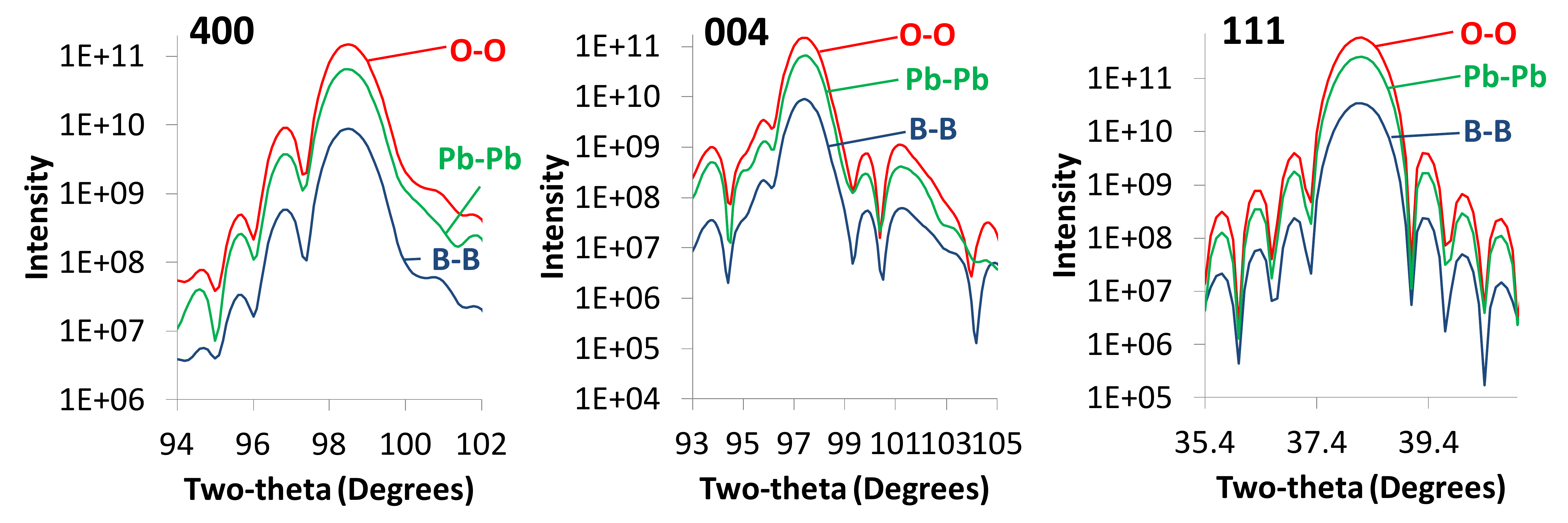}
\caption{\label{Satellites_x51}The contribution of the Pb-, O- and $B$-sublattices to the neutron scattering intensity of the $400$, $004$, and $111$ 
reflections of the $x=0.51$ cluster. Logarithmic scale is used in order to show the weak satellite peaks more clearly. }
\end{center}
\end{figure}

Correlated displacement and substitutional disorder is most clearly seen in the $h00$ and $00l$ type reflections, whereas the $hhh$ reflections possess nearly symmetric distribution of satellite peaks, 
see Fig. \ref{Satellites_x51}. The remarks found in the case of periodical composition and displacement disorder (see Table \ref{Defects}) are seen in Figs. \ref{Satellites} and \ref{Satellites_x51}: The intensities of 
the satellite peaks belonging to the same pair centred at each Bragg reflection are not identical.  The signatures are generic to PZT nanoclusters, though the details depend on the cluster statistics. 

\subsection{\label{ellipsoid_section}Ellipsoid shaped single domain clusters}
\begin{figure}[h!]
\begin{center}
\includegraphics[angle=0,width=12cm]{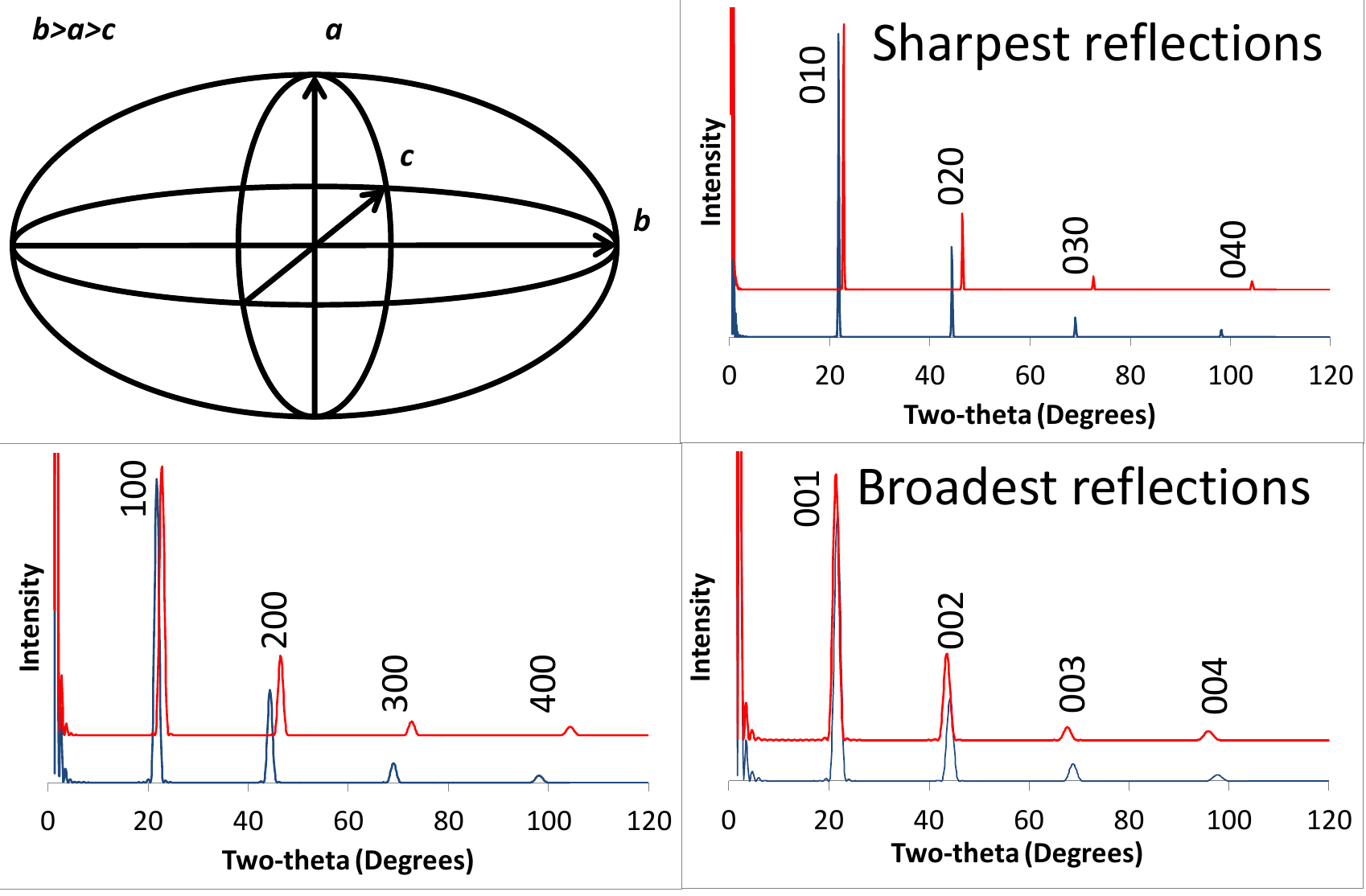}
\caption{\label{Ellipsoid} Scattering intensities from an ellipsoid shaped PbTiO$_3$ (red line) and a Pb(Zr$_{0.52}$Ti$_{0.48}$)O$_3$ (blue line) cluster. 
The ellipsoid axes  $a$, $b$ and $c$ were 24, 108 and 18 cell edges long, respectively ($N_x=24$, $N_y=108$ and $N_z=18$) 
and were parallel to the $a_x$, $a_y$ and $a_z$ axes.}
\end{center}
\end{figure}

Figure \ref{Ellipsoid} shows the x-ray scattering profiles along three directions for $x=0$ and $\approx 0.52$ clusters.
Table \ref{EllipsoidParameters} gives the average lattice parameter and cluster size estimates, obtained through Bragg 
and Scherrer equations, respectively. The widths correspond to the values estimated from the Scherrer equation, though 
they are systematically smaller than the cluster dimensions, being about 96 \% from the maximum ellipsoid dimension. 
The error is very similar to the error obtained in the case 
of spherical clusters for which the Scherrer equation (\ref{Scherrer}) holds. As in the case of spherical clusters the Bragg equation values are 
rather close to the known average values. Partially the error is due to the reasons discussed in the context of spherical clusters.
The impact of the width of the scattering intensity on the average lattice parameter 
estimation is rather small.

\begin{table}[h!]
\begin{center}
\caption{\label{EllipsoidParameters}Cluster metrics and the values as obtained through Bragg equation ($a_t$, $b_t$ and $c_t$) and Scherrer 
equation ($a$, $b$ and $c$). Values (all in units of \AA) are given in pairs, the first is the cluster (accurate) value, the second one is obtained through the 
Bragg/Scherrer equation.}
\begin{tabular}{l l l l}
\hline
$x$           & $a_t$                             & $b_t$                               & $c_t$                                         \\
0               & 3.900000, 3.900300     & 3.900000, 3.900015        & 4.150000, 4.150105                 \\
0.514485 & 4.079813, 4.078121      & 4.074603, 4.074910       & 4.090201, 4.092251                 \\
                 & $a$                                & $b$                                   & $c$                                            \\
0               & 93.600000, 89.503502 & 421.200000, 405.758631 & 74.700000, 71.396015            \\
0.514485 & 97.700121, 93.289568  & 439.059234, 422.897821 & 73.485345, 70.458609           \\
\hline
\end{tabular}
\end{center}
\end{table}

\subsection{\label{DomainSection}Domains in spherical clusters}
\begin{figure}[h!]
\begin{center}
\includegraphics[angle=0,width=10cm]{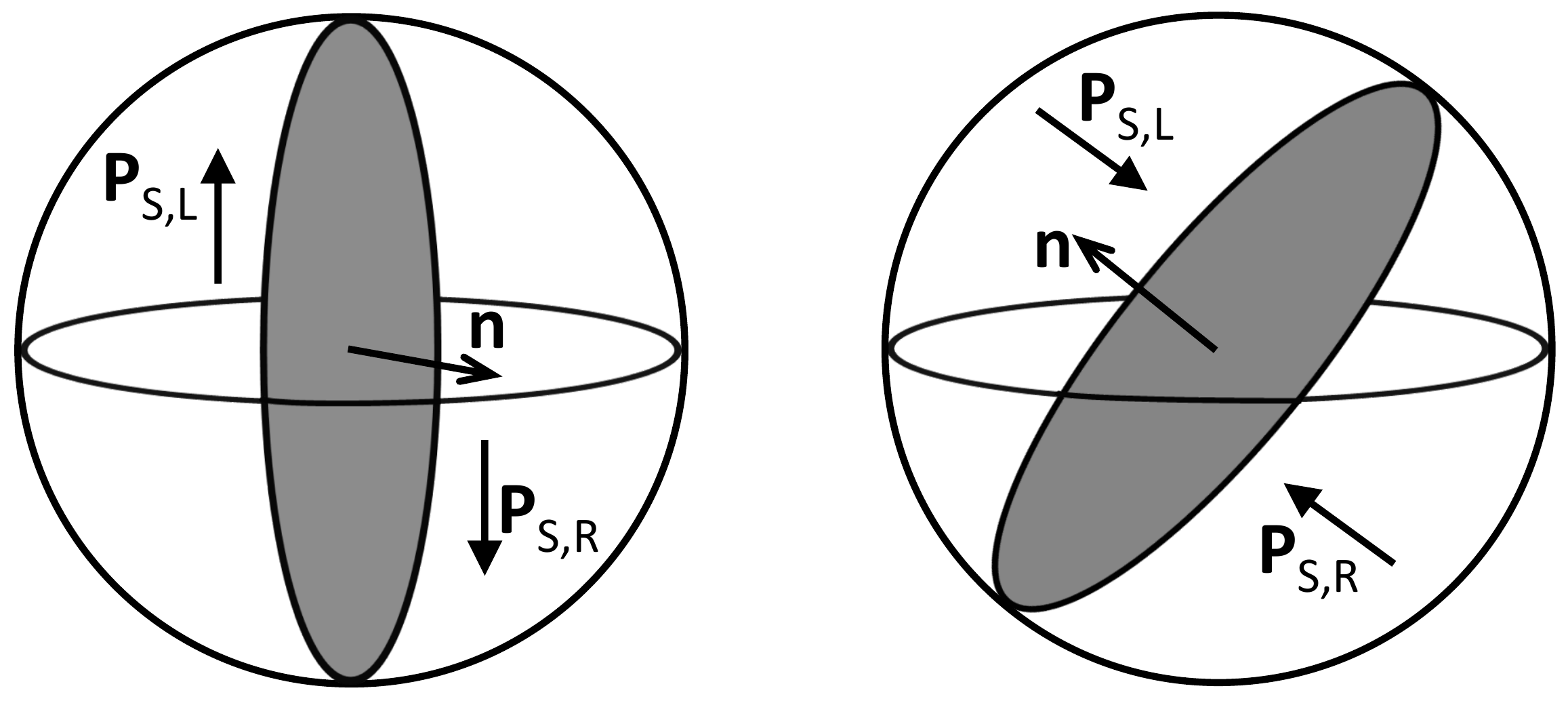}
\caption{\label{Domains} $180^{\circ}$-domain walls (grey) in titanium-rich (left-hand side, termed T-domains) and zirconium-rich PZT (right-hand side, termed R-domains). Spontaneous 
polarization directions on the left- and right-hand side and the domain wall normal are indicated by $\mathbf{P}_{\mathrm{S,L}}$, 
$\mathbf{P}_{\mathrm{S,R}}$ and $\mathbf{n}$, respectively.  }
\end{center}
\end{figure}

\begin{figure}[!p]
\begin{center}
\includegraphics[angle=0,width=13cm]{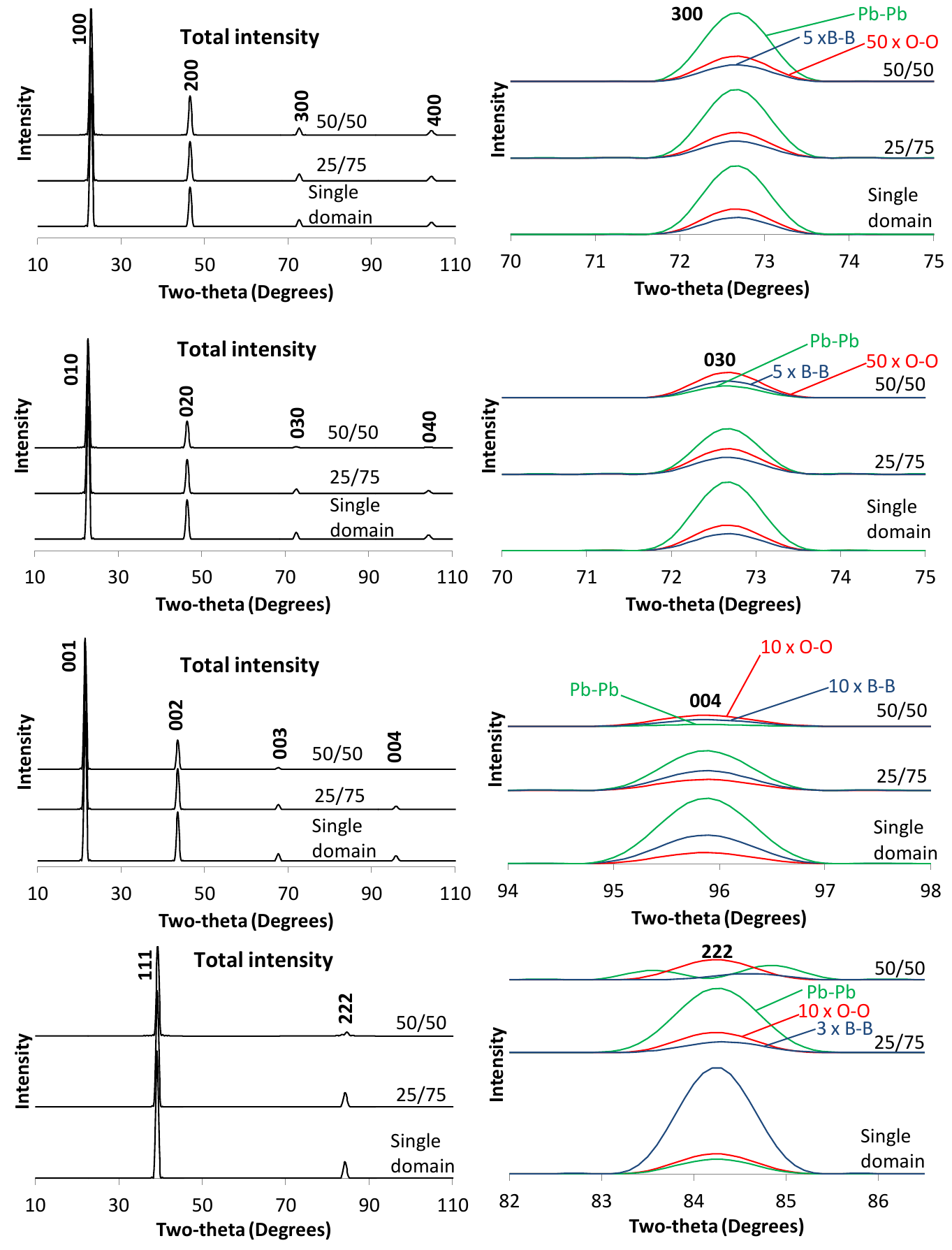}
\caption{\label{Domain_PT} Selected x-ray scattering peaks computed for 
(wavelength 1.540562 \AA) two T-domain wall positions (volume ratios 50/50 and 25/75) and a single domain PbTiO$_3$ cluster. 
The green curves give the contribution due to the Pb-sublattice, red give the oxygen sublattice contribution, multiplied by a factor of 50 (300 and 030 reflections) or 10 (004 and 222 reflections), and blue give the 
$B$-cation contribution multiplied by a factor of 5 (300 and 030 reflections), 10 (004 reflection) or 3 (222 reflection). Total intensities (which include also Pb-O, Pb-$B$ and alike terms) are given in the left-hand column.}
\end{center}
\end{figure}

\begin{figure}[!p]
\begin{center}
\includegraphics[angle=0,width=13cm]{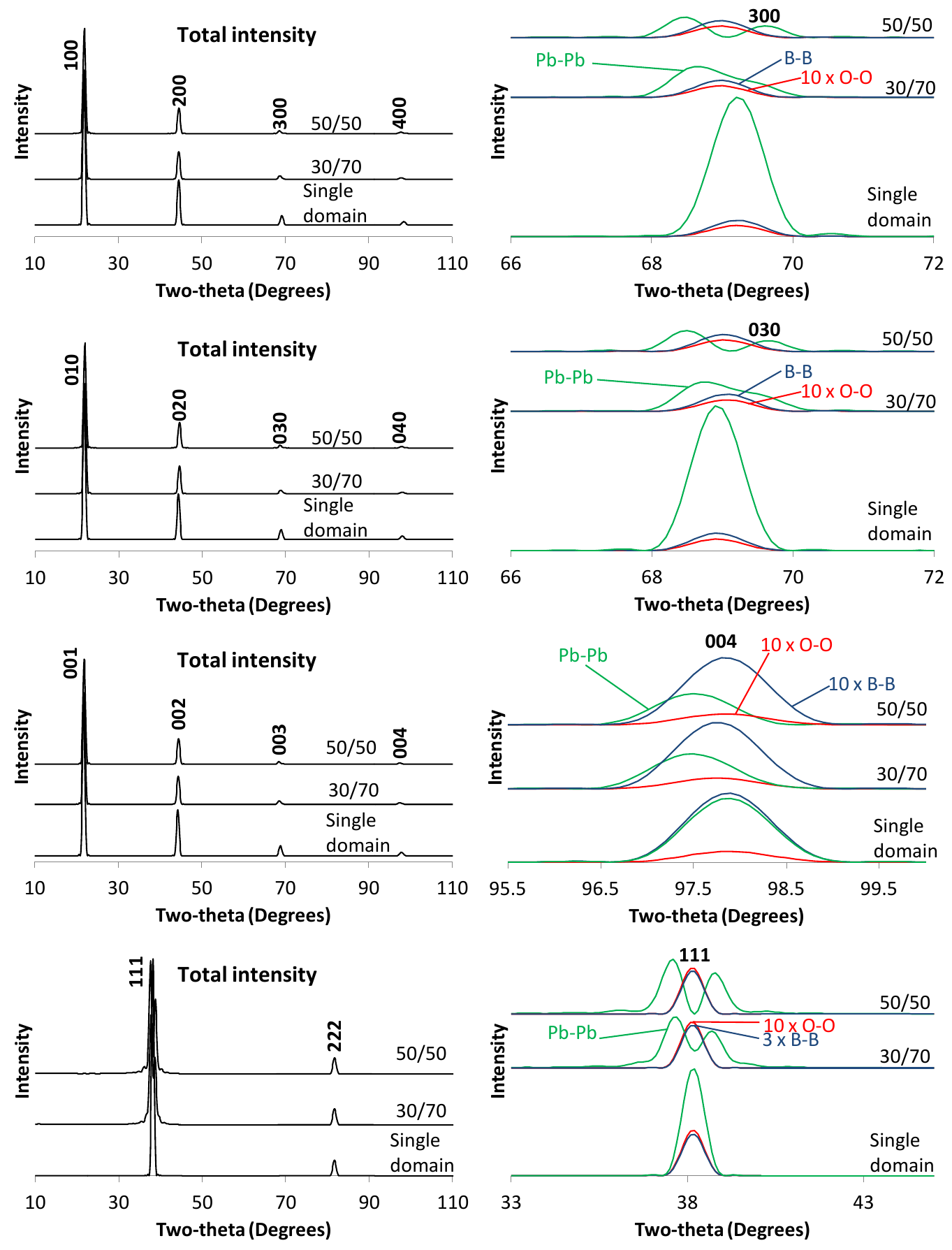}
\caption{\label{Domain_PZT} Selected x-ray scattering peaks computed for 
(wavelength 1.540562 \AA) two R-domain wall positions (volume ratios 50/50 and 30/70) and a single domain  Pb(Zr$_{0.52}$Ti$_{0.48}$)TiO$_3$ cluster. 
The green curves give the contribution due to the Pb-sublattice, red give the oxygen sublattice contribution and blue give the 
$B$-cation contribution. To show the features more clearly, the intensities of the oxygen lattice is multiplied by 10 and  the and $B$-cation lattice by a 
factor of 1 ($300$ and $030$ reflections), 3 ($111$ reflection) or 10 ($004$ reflection). Total intensities (which include also Pb-O, Pb-$B$ and alike terms) 
are given in the left-hand column.}
\end{center}
\end{figure}

As an example of a complex defect system $180^{\circ}$ domain walls and domains in spherical clusters are modelled. Fig. \ref{Domains} 
illustrates the structural model. 
The cases treated below possess complex combination of displacement disorder (when the boundary halves the particle, left-hand side domain can be obtained from the right-hand side domain by atomic 
displacements) and substitutional disorder.
The cluster for simulations is constructed as explained in section \ref{FlowChart}, 
except that the cation displacements directions are reversed at different domains. In each domain the cation positions 
are relaxed so that the bond-valence sum criteria is fulfilled. 
In contrast to many standard models, the two domains are not considered separately; instead 
the scattering intensity is computed for the \emph{entire cluster}. The purpose of the simulation is to show that 
(i) domain structure has a significant impact on the scattering intensity, (ii) the method suits for \emph{in-situ} modelling (e.g., 
polarization reversal studies), (iii) the impact of different elements can be isolated and (iv) certain reflections can no more be described 
by a single asymmetric peak, even if they would originally correspond to a single symmetric peak. 
Two domains types, referred to as T- and R-domains, are considered. In the T-domain the domain 
wall is perpendicular to the $\langle 100 \rangle$ direction and in the case of the R-domain the wall is perpendicular to the 
$\langle 111 \rangle$ direction.
The presence of T- and R-domains is most evidently revealed as a split of the $hhh$-type reflections. The split itself depends on the 
domain wall position in the cluster, as is seen from Figs. \ref{Domain_PT} and \ref{Domain_PZT}. In the case of R-domain and $111$ reflection 
also the scattering intensity in the tail regions of the Bragg peak is considerable, consistently with the known strong diffuse scattering 
observed in many Pb-perovskites. As  reflection $111$ indicates, see Fig. \ref{Domain_PZT}, the strong diffuse scattering is due to the 
Pb-Pb scattering and is easily seen when the particle is divided into domains. 
The most significant contribution to the split is from 
Pb-ions, which are displaced from the corners to a cell diagonal direction. The parallel epipeds are not identical and 
thus there is a variation in the Pb-displacement directions within a domain. 
In T-domains the $h00$ type reflections are left intact by the domain wall, whereas the intensity depends strongly on the domain wall position in 
the case of the $0k0$ and $00l$ reflections. In R-domains all reflections intensities and line shapes strongly depend on the 
domain wall position.

\paragraph{Applications.}By adjusting domain wall position and orientation and cation displacement 
direction different cases (e.g., 71$^{\circ}$ domain walls) can be constructed in a straightforward manner. 
Domain wall motion takes place in polarization ($\mathbf{P}$) 
switching. 
Also structural changes occurring in the domain wall region can be implemented into the model by constructing a domain wall 
with a finite thickness. 
T-domain boundary addressed above fulﬁls the 
electrical boundary condition, $\mathbf{\nabla} \cdot \mathbf{P}=0$, so that the boundary is not charged. The mechanical 
compatibility condition is satisfied as there is no abrupt change in the average lattice parameters across the boundary (i.e., the 
boundary is stress-free). Also more complex cases, such as the formation of impurity phase or cases not fulfilling the electrical 
and mechanical boundary conditions (head-to-head R-domain is an example) can be constructed. A plausible application would be a modelling of \emph{in-situ} measurements 
of domain walls in ferroelectric materials. Due to the high brightness x-ray synchrotron radiation based experimental techniques 
are evident methods for addressing time-dependent phenomena. As an example, the nonlinear effects in the coupling of polarization 
with elastic strain and the initial stage of polarization switching were addressed in refs. \cite{Grigoriev_2008,Grigoriev_2009}. In 
these studies capacitors containing 35 nm thick epitaxial Pb(Zr$_{0.20}$Ti$_{0.80}$)O$_3$ ferroelectric thin films were studied 
by time-resolved x-ray microdiffraction technique in which high-electric field up to several hundred MV/m pulses were synchronized 
with the synchrotron x-ray pulses. Laboratory scale measurements can also be used to address the \emph{in situ} structural 
changes in thin films. An example is given in ref. \cite{Biegalski_2014} which reports the changes in lattice parameter (chemical expansivity)
and its further use for quantifying oxygen reduction reaction processes and vacancy concentration changes in  
La$_{0.80}$Sr$_{0.20}$CoO$_{3-\delta}$ thin films under chemical and voltage stimuli. 

An example of the use of an laboratory x-ray diffractometer to address time-dependent ferroelectric domain reversal is given in ref. \cite{Pramanick}, 
where the changes in the volume fractions of the 90$^{\circ}$ domains parallel to the electric field direction were calculated from the intensities of 
the $\{002\}$ diffraction peaks. 

Also magnetic scattering can be treated in a manner analogous to the nuclei scattering. In magnetic materials the domain 
boundary region often have spatially large extent (e.g., N\'eel and Bloch walls) also in a direction perpendicular to the domain wall.
In multiferroic materials the ferroic properties are not necessarily well coupled, and thus it is crucial to be able to isolate different 
contributions to the scattering intensity. For instance, magnetization reversal may not be accompanied by apparent changes in 
nuclei arrangements.

All above disorder cases can be combined to model the often complex structures of multilayer thin films. Especially in structures formed from 
very thin layers (say, of the order of few nanometers) the scattering intensity should not be treated as originating from independent 
layers. A better way is to model the entire structure and to impose the required boundary conditions for interfaces. 

\section*{Conclusions}
Applications of x-ray and neutron scattering techniques for analysing defects were reviewed. Focus was on the common approaches 
applied for modelling defects. 
A method for analysing scattering data collected on nanoparticles was described with necessary compatibility conditions. The 
method is capable of isolating different contributions to the scattering intensity, such as element specific scattering, microstrain in 
solid-solutions, particle size and shape effects and domains: structural disorder is not averaged away. Scattering measurements 
provide information without destructive sample preparation. A case study on lead-zirconate-titanate nanoparticles was given. 
Potential applications include nanoparticles, disordered nanoparticles and in-situ studies of structural changes in ferroic materials.

\section*{Acknowledgments}
CSC - IT Center for Science Ltd. is acknowledged for providing a computing environment.


\newpage
\pagestyle{myheadings}
\markright{Appendix I}

\section{Instrument specific corrections}
Scattering geometry and the nature of the incident radiation (for instance, x-rays or polarized or unpolarized neutron beams) affects the 
observed scattering intensity. The simplest correction is the diffraction geometry dependent Lorentz factor $L$, a trigonometric factor 
due to the fact that different reflections stay different times at the reflection position. In the case of a constant angular speed 
and single crystal sample $L$ is $1/\sin2\theta$ (see, e.g. ref. \cite{Massa}). For powders the prevailing measurement geometry is the 
Bragg-Brentano, or $\theta/2\theta$ geometry, in which the sample is rotated at constant speed. If one assumes that the incident 
beam probes large set of randomly oriented crystals a further geometrical factor $1/\sin\theta$ is required as only a fraction of the 
crystals are in a reflection position with respect to the detector. The Lorentz factor is the same for x-rays and neutrons. Thus, the 
neutron scattering intensities computed through Eq.(\ref{IN}) were multiplied by a factor
\begin{center}
\begin{equation}\label{LN}
1/(2\sin^2\theta\cos\theta)
\end{equation}
\end{center}
In the case of x-rays one needs a further correction to take the polarization of x-rays into account. Typically x-rays emerging from a 
x-ray tube are unpolarized, and after reflecting from the diffracting plane the component of the electric field parallel to the plane is not 
attenuated, whereas the electric field component perpendicular to the diffracting plane is attenuated by a factor of $\cos^2\theta$, so that the 
total intensity is reduced by a polarization factor $p=(1+\cos^2\theta)/2$. Monochromators affect the polarization factor and modified expressions 
for $p$ are required: $p=(1+K\cos^2\theta)/(1+K)$ when a monochromator for an incident beam is used \cite{Massa}. $K$ is a constant typically 
close to unity.

The effect of Lorentz and polarization corrections is typically put together and expressed as the $Lp$ factor, which multiplies the scattering power 
or, in the case of crystals, $\vert F_{hkl}\vert^2$. In this study the x-ray scattering intensities were multiplied by a factor
\begin{center}
\begin{equation}\label{Lp}
(1+\cos^22\theta)/(\sin^2\theta\cos\theta)
\end{equation}
\end{center}

\section{Line shapes in diffraction experiments}
Instrument also affects the observed lineshape. When the scattering power is computed by equation (\ref{IN}) as a function of $2\theta$ for a chosen direction of $\mathbf{s}$, the cluster size and shape dictates the linewidths 
and, if the cluster is small, is almost entirely responsible for the broadening.
In contrast, when the intensity for a specific reflection is computed as $\vert F_{hkl}\vert^2$, a specific peak profile must be assumed. Typically, 
one needs to consider the sample size and shape, possible defects causing symmetric and asymmetric broadening, and the instrumental contribution 
as is done in conventional Rietveld refinement. The functions describing different contributions are convoluted, which is computationally heavy task.
Correspondingly, a large number of different types of profile functions have been introduced to model different cases.

\end{document}